\documentclass[12pt]{iopart}
\usepackage{graphicx}
\eqnobysec

\def\numpartsappendix{\addtocounter{equation}{1}%
     \setcounter{eqnval}{\value{equation}}%
     \setcounter{equation}{0}%
     \def\theequation{\ifnumbysec
     \Alph{section}.\arabic{eqnval}{\it\alph{equation}}%
     \else\Alph{eqnval}{\it\alph{equation}}\fi}}
\def\endnumpartsappendix{\def\theequation{\ifnumbysec
     \Alph{section}.\arabic{equation}\else
     \Alph{equation}\fi}%
     \setcounter{equation}{\value{eqnval}}}

\begin{document}


\title[High-energy asymptotic expansion of the Green function]
{High-energy asymptotic expansion of the Green function
for one-dimensional Fokker-Planck and Schr\"odinger equations}

\author{Toru Miyazawa}

\address{Department of Physics, Gakushuin University, 
Tokyo 171-8588, Japan}
\ead{toru.miyazawa@gakushuin.ac.jp}
\begin{abstract}
A new formalism is presented for high-energy analysis of the Green function for Fokker-Planck and Schr\"odinger equations in one dimension. 
Formulas for the asymptotic expansion in powers of the inverse wave number are derived, and conditions for the validity of the expansion are studied through the analysis of the remainder term. This method is applicable to a large class of potentials, including the cases where the potential $V(x)$ is infinite as $x \to \pm \infty$.
The short-time expansion of the Green function is also discussed.
\end{abstract}

\pacs{03.65.Nk, 02.30.Hq, 02.50.Ey}
\maketitle


\section{Introduction}
The one-dimensional diffusion in an external potential $V(x)$ is described by the Fokker-Planck equation \cite{risken}
\begin{equation}
\label{1eq1.1}
\frac{\partial}{\partial t}P(x,t)
=\frac{\partial^2}{\partial x^2}P(x,t)-2\frac{\partial}{\partial x}[f(x)P(x,t)],
\end{equation}
where $f(x)$ is related to the potential $V(x)$ by
\begin{equation}
\label{1eq1.2}
f(x)=-\frac{1}{2}\frac{d}{dx}V(x).
\end{equation}
With $P(x,t)\equiv e^{-k^2 t} \phi(x)$, equation (\ref{1eq1.1}) reduces to the time-independent form
\begin{equation}
\label{1eq1.3}
-\frac{d^2}{dx^2}\phi(x)+2\frac{d}{dx}[f(x)\phi(x)]=k^2\phi(x).
\end{equation}
In addition to being a fundamental equation for nonequilibrium phenomena, the Fokker-Planck equation is of particular importance in its relation to the Schr\"odinger equation.
Setting $\phi(x) \equiv e^{-V(x)/2}\psi(x)$, we can transform (\ref{1eq1.3}) into a steady-state Schr\"odinger equation
\begin{equation}
\label{1eq1.4}
-\frac{d^2}{dx^2}\psi(x)+V_{\rm S}(x)\psi(x)=E\psi(x),
\end{equation}
where $E=k^2$, and
\begin{equation}
\label{1eq1.5}
V_{\rm S}(x)=f^2(x)+f'(x).
\end{equation}
Historically, methods developed for the Schr\"odinger equation have been applied to the study of the Fokker-Planck equation. But this should be the other way around as well.
Theoretically, the Fokker-Planck equation is often more convenient to deal with than the Schr\"odinger equation. With the Fokker-Planck equation we can carry out a more systematic analysis, and the results obtained for the Fokker-Planck equation can be applied to quantum-mechanical problems described by the Schr\"odinger equation.
In this paper, we study the high-energy (large-$\vert k \vert$) behavior of the Green function for (\ref{1eq1.3}) or, equivalently,  (\ref{1eq1.4}). 
Although we shall mainly work with the Fokker-Planck equation, the results of this paper are directly applicable to the Schr\"odinger equation, too.

One-dimensional quantum scattering has a long history. 
In particular, high-energy asymptotic behavior of the Green function and related functions have been studied over the decades by both physicists and mathematicians.
This classical topic has recently attracted attention in connection with inverse problems and the theory of integrable systems, and this area of research remains active even to the present day \cite{newton}--\cite{rybkin2}.

An essential matter in the study of an asymptotic expansion is the estimation of the remainder term. 
In conventional methods, which are mostly based on an integral equation, 
it is necessary to impose strong conditions on the potential in order to control the remainder term. For example, the integrability of $V_{\rm S}$ is often required. For non-integrable $V_{\rm S}$, different specific methods need to be used.  
There has not yet been a formalism in which the asymptotic analysis of the Green function can be carried out systematically and in a unified way for a variety of potentials that may not necessarily be integrable and that may not even be finite as $\vert x \vert \to \infty$. 
In this paper, we take a totally new approach to this problem, and present a new method for the high-energy analysis of the Green function which is applicable to a larger class of potentials.

Our method is based on the analysis of reflection coefficients. 
Reflection coefficients are fundamental quantities in scattering theory, and they serve as building blocks for constructing the Green function. The Green function can be expressed solely in terms of reflection coefficients for semi-infinite intervals \cite{expressions}. 
High-energy behavior of the reflection coefficients was studied in a previous paper \cite{analysis}, and formulas for their asymptotic expansion were derived there. In the present paper, we apply these results to the Green function, and derive the expansion in powers of the inverse wave number. 
The coefficients of the expansion are expressed in a simple form in terms of a linear operator, and the remainder term is expressed in terms of transmission and reflection coefficients for finite intervals.
The validity of the asymptotic expansion can be studied by using this expression for the remainder term. The short-time expansion of the Green function can also be obtained by this method.

We assume that the (Fokker-Planck) potential $V(x)$ is a real function which is finite for finite $x$, and which either converges to a finite limit or diverges to infinity ($+ \infty$ or $-\infty$) as $x \to \pm\infty$. 
The quantity $k$ is taken to be a complex number with ${\rm Im}\,k\geq 0$.

In sections~2 and 3, we review the relevant results of previous papers. 
We derive the asymptotic expansion of the Green function in sections~4--6, and discuss the conditions for its validity in sections~7 and 8. In section~9, the short-time expansion is studied. In section~10, we explain how our method can be applied to the Schr\"odinger equation. Various examples are given in section~11.

\section{Reflection coefficients and the Green function}

We define the time-dependent Green function $G_{\rm F}(x,y;t)$ as the solution of
\begin{equation}
\label{1eq2.1}
\left[\frac{\partial}{\partial t}
-\frac{\partial^2}{\partial x^2}+
2\frac{\partial}{\partial x}f(x)\right]G_{\rm F}(x,y;t)=\delta(x-y)\delta(t)
\end{equation}
with the condition that $G_{\rm F}(x,y;t)=0$ for $t<0$. 
Physically, $G_{\rm F}(x,y;t)$ is the probability density of finding the diffusing particle at position $x$ at time $t$, under the condition that it was initially at position $y$.
We also define its Fourier transform
\begin{equation}
\label{1eq2.2}
G_{\rm F}(x,y;\omega)\equiv \int_0^\infty e^{i\omega t}G_{\rm F}(x,y;t)\,dt.
\end{equation}
This $G_{\rm F}(x,y;\omega)$ is the Green function for equation (\ref{1eq1.3}) with $k^2=i\omega$. 
In the same way, we define the retarded Green function for the Schr\"odinger equation as the solution of
\begin{equation}
\label{1eq2.3}
\left[i\frac{\partial}{\partial t}+\frac{\partial^2}{\partial x^2}
-V_{\rm S}(x)\right]G_{\rm S}(x,y;t)=\delta(x-y)\delta(t)
\end{equation}
satisfying the condition $G_{\rm S}(x,y;t)=0$ for $t<0$. 
Its Fourier transform defined by
\begin{equation}
\label{1eq2.4}
G_{\rm S}(x,y;E) \equiv \int_0^\infty e^{iEt}G_{\rm S}(x,y;t)\,dt
\end{equation}
is the Green function for the steady-state Schr\"odinger equation (\ref{1eq1.4}), satisfying
\begin{equation}
\label{1eq2.5}
\left[\frac{\partial^2}{\partial x^2}-V_{\rm S}(x) + E\right]G_{\rm S}(x,y;E)
=\delta(x-y).
\end{equation}
It is easy to see that $G_{\rm F}$ and $G_{\rm S}$ are related by
\begin{equation}
\label{1eq2.6}
G_{\rm F}(x,y;\omega)=-e^{-[V(x)-V(y)]/2}G_{\rm S}(x,y;E=i\omega).
\end{equation}
For convenience, we define
\begin{equation}
\label{1eq2.7}
G(x,y;k)\equiv 2ik G_{\rm S}(x,y;E=k^2)
\end{equation}
as a function of complex $k$, and deal with this $G$ instead of $G_{\rm F}$ or $G_{\rm S}$. (We shall refer to $G$ as the Green function, too.)
Since $G(x,y;k)=G(y,x;k)$, without loss of generality we assume that $x \geq y$.

To introduce our basic expression for the Green function, let us first define the transmission and reflection coefficients for finite intervals.
We consider an interval $(x_1,x_2)$, and define
\begin{equation}
\label{1eq2.8}
\bar V(x)\equiv
\cases{
V(x_1) & $(x<x_1)$ \\
V(x) & $(x_1\leq x \leq x_2)$ \\
V(x_2) & $(x_2<x)$
},
\qquad 
\bar f(x)\equiv -\frac{1}{2}\frac{d}{dx}\bar V(x).
\end{equation}
Namely, $\bar V(x)$ is identical to $V(x)$ within $(x_1,x_2)$  and constant outside this interval. We consider equation (\ref{1eq1.3}) with $f(x)$ replaced by $\bar f(x)$:
\begin{equation}
\label{1eq2.9}
-\frac{d^2}{dx^2}\phi(x)+2\frac{d}{dx}[\bar f(x)\phi(x)]=k^2\phi(x).
\end{equation}
(In general, the left-hand side of this equation contains delta functions at $x=x_1$ and $x=x_2$ coming from the derivative of $\bar f$.) 
Since $\bar f(x)=0$ for $x<x_1$ and $x>x_2$, equation (\ref{1eq2.9}) has two solutions of the form
\refstepcounter{equation}
\label{1eq2.10}
\addtocounter{equation}{-1}
\numparts
\begin{equation}
\phi_1(x)=
\cases{
e^{[V(x_2)-V(x_1)]/2} \tau(x_2,x_1;k) e^{-ik(x-x_1)} & $x<x_1$, \\
e^{-ik(x-x_2)} + R_r(x_2,x_1;k) e^{ik(x-x_2)} & $x>x_2$, \\
}
\end{equation}
\begin{equation}
\phi_2(x)=
\cases{
e^{ik(x-x_1)} + R_l(x_2,x_1;k) e^{-ik(x-x_1)} &$x<x_1$, \\
e^{-[V(x_2)-V(x_1)]/2} \tau(x_2,x_1;k) e^{ik(x-x_2)} &  $x>x_2$. \\
}
\end{equation}
\endnumparts
(The factor $e^{\pm [V(x_2)-V(x_1)]/2}$ in front of $\tau$ is necessary since $\phi_1$ and $\phi_2$ are solutions of the Fokker-Planck equation, not the Schr\"odinger equation.)
The transmission coefficient $\tau$, the right reflection coefficient $R_r$, and the left reflection coefficient $R_l$ for the interval $(x_1,x_2)$ are defied by (\ref{1eq2.10}).

We can let $x_1 \to -\infty$ in (\ref{1eq2.10}a) or $x_2 \to +\infty$ in (\ref{1eq2.10}b) to define the reflection coefficients for semi-infinte intervals:
\begin{equation}
\label{1eq2.11}
\fl
R_r(x_2,-\infty;k)=\lim_{x_1 \to -\infty} R_r(x_2,x_1;k), \qquad
R_l(\infty,x_1;k)=\lim_{x_2 \to +\infty}R_l(x_2,x_1;k).
\end{equation}
Let us define
\begin{equation}
\label{1eq2.12}
S_r(x,k) \equiv \frac{R_r(x,-\infty;k)}{1+R_r(x,-\infty;k)}, 
\qquad
S_l(x,k) \equiv \frac{R_l(\infty;x;k)}{1+R_l(\infty,x;k)},
\end{equation}
and
\begin{equation}
\label{1eq2.13}
S(x,k)\equiv S_r(x,k)+S_l(x,k).
\end{equation}
It was shown in \cite{expressions} that the Green function can be expressed in terms of this $S$ as
\begin{equation}
\label{1eq2.14}
\fl
G(x,y;k)=
\frac{1}{\sqrt{[1-S(x,k)][1-S(y,k)]}}
\exp\left[
ik(x-y)-ik\int_y^x S(z,k)\,dz
\right].
\end{equation}
 The logarithm of (\ref{1eq2.14}) reads
\begin{eqnarray}
\label{1eq2.15}
\fl
\log G(x,y;k)=ik(x-y)-ik\int_y^x S(z,k)\,dz -\frac{1}{2}\log[1-S(x,k)]-\frac{1}{2}\log[1-S(y,k)].
\nonumber \\
\end{eqnarray}
We shall carry out the analysis of the Green function on the basis of this expression.

\section{High-energy expansion formula for the reflection coefficients}

In this section, we review the formulas derived in \cite{analysis} for the high-energy expansion of the reflection coefficients. Here we deal only with $R_r$. (Corresponding formulas for $R_l$ can be obtained in the same way.)
First, we define the generalized transmission and reflection coefficients, with an additional argument $\xi$, as
\refstepcounter{equation}
\label{1eq3.1}
\addtocounter{equation}{-1}
\numparts
\begin{equation}
\bar R_r(x,y;\xi;k) \equiv \frac{R_r(x,y;k)-\xi}{1-\xi R_r(x,y;k)},
\end{equation}
\begin{equation}
\bar R_l(x,y;\xi;k) \equiv 
R_l(x,y;\xi;k) + \frac{\xi \tau^2(x,y;\xi;k)}{1-\xi R_r(x,y,\xi;k)},
\end{equation}
\begin{equation}
\bar \tau(x,y;\xi;k)\equiv 
\frac{\sqrt{1-\xi^2}\, \tau(x,y;k)}{1-\xi R_r(x,y;k)}.
\end{equation}
\endnumparts
We also define the operator ${\cal M}$ which acts on functions of $x$ and $\xi$ as
\begin{equation}
\label{1eq3.2}
\fl
{\cal M} g(x,\xi) \equiv
-f(x)\left[\left(\xi-\xi^{-1}\right)g(x,\xi)+\xi^{-1}g(x,0)\right]
+\xi^{-1}\int_0^\xi d\xi \,\frac{\partial}{\partial x}g(x,\xi),
\end{equation}
where $g(x,\xi)$ is an arbitrary function, and $f(x)$ is the function in (\ref{1eq1.1})--(\ref{1eq1.3}).

It was shown in \cite{analysis} that, for an arbitrary nonnegative integer $N$,  
\begin{eqnarray}
\label{1eq3.3}
\fl
&\bar R_r(x,-\infty;\xi;k)
\nonumber 
\\
\fl
&\qquad= -\xi + \frac{1}{2ik} \bar c_1(x,\xi) + \frac{1}{(2ik)^2}\bar c_2(x,\xi) + \cdots
+ \frac{1}{(2ik)^N}\bar c_N(x,\xi)+ \bar \delta_N(x,\xi,k),
\end{eqnarray}
where
\begin{equation}
\label{1eq3.4}
\bar c_n(x,\xi)=-(1-\xi^2) {\cal M}^{n-1} f(x),
\end{equation}
\begin{equation}
\label{1eq3.5}
\bar \delta_N(x,\xi;k)
=\frac{1}{(2ik)^N}\int_{-\infty}^x\bar \tau^2(x,z;\xi;k) K_N(z,\bar R_l(x,z;\xi;k))\,dz,
\end{equation}
\begin{equation}
\label{1eq3.6}
K_n(x,\xi) \equiv -\left(1+ \xi \frac{\partial}{\partial \xi}\right)
\frac{1}{1-\xi^2}\bar c_{n+1}(x,\xi).
\end{equation}
In (\ref{1eq3.5}), $K_N(z,\bar R_l(x,z;\xi;k))$  is the quantity obtained from $K_N(z,\xi)$ by the substitution $\xi \to \bar R_l(x,z;\xi;k)$.
It is convenient to define
\begin{equation}
\label{1eq3.7}
\tilde c_n(x,\xi) \equiv \frac{1}{1-\xi^2} \bar c_n(x,\xi), 
\end{equation}
so that (\ref{1eq3.4}) and (\ref{1eq3.6}) read
\begin{equation}
\label{1eq3.8}
\tilde c_n(x,\xi)=-{\cal M}^{n-1} f(x),
\end{equation}
\begin{equation}
\label{1eq3.9}
K_n(x,\xi) = -\left(1+ \xi \frac{\partial}{\partial \xi}\right)
\tilde c_{n+1}(x,\xi).
\end{equation}
It is easy to calculate $\tilde c_n$ from (\ref{1eq3.8}) by using definition (\ref{1eq3.2}). 
We have
\begin{eqnarray}
\label{1eq3.10}
\fl
{\tilde c}_1=-f, \qquad
{\tilde c}_2=-f'+f^2 \xi, \qquad
{\tilde c}_3=-f''+f^3+2ff'\xi - f^3 \xi^2,
\nonumber\\
\fl
{\tilde c}_4=-f'''+5f^2f'-(2f^4-{f'}^2-2ff'')\xi -3f^2f'\xi^2 +f^4\xi^3,
 \quad {\rm etc.}
\end{eqnarray}
Obviously $\tilde c_n$ is an $(n-1)$th degree polynomial in $\xi$, whose coefficients consist of the powers of $f$ and its derivatives. 
The $K_n$'s are obtained from (\ref{1eq3.9}) and (\ref{1eq3.10}) as
\begin{eqnarray}
\label{1eq3.11}
\fl
K_0=f, \qquad
K_1=f'-2f^2 \xi, \qquad
K_2=f''-f^3-4ff'\xi +3 f^3 \xi^2,
\nonumber \\
\fl
K_3=f'''-5f^2f'+2(2f^4-{f'}^2-2ff'')\xi +9f^2f'\xi^2 -4f^4\xi^3,
\quad {\rm etc.}
\end{eqnarray}

Equations (\ref{1eq3.3})--(\ref{1eq3.6}) hold for any $f(x)$ as long as $\bar c_n$ and $\bar \delta_N$ make sense.  However, (\ref{1eq3.3}) is meaningful as a high-energy expression only if
\begin{equation}
\label{1eq3.12}
\lim_{\vert k \vert \to \infty}k^N \bar \delta_N(x,\xi,k)=0.
\end{equation}
Using (\ref{1eq3.5}) with (\ref{1eq3.9}), 
we can study the conditions for (\ref{1eq3.12}) to hold.
For simplicity, let us assume that $f(x)$ and all its derivatives are monotone for sufficiently large $\vert x \vert $.
(This condition is unnecessarily strong and can be relaxed, but we make this assumption in order to simplify the presentation.)
Then it can be shown\footnote{
The proof is given in~\cite{analysis} only for $\xi=0$, but this is sufficient.
(See the comment below equation (5.16) of~\cite{analysis}.) 
If (\ref{1eq3.12}) holds for $\xi=0$ then it holds for any $\xi$. This can be easily shown by substituting the expansion of $R_r$ (equation (\ref{1eq3.3}) with $\xi=0$) into the right-hand side of (\ref{1eq3.1}a). 
} 
that:

\begin{enumerate}
\item
When $\vert k \vert \to \infty$ with fixed $\arg k$ in the range $0<\arg k<\pi$, 
equation (\ref{1eq3.12}) holds if ($f$ is $(N-1)$ times differentiable and) $f^{(N-1)}$ is continuous and  piecewise smooth\footnote{
This last condition is described as ^^ ^^ piecewise differentiable" in \cite{analysis}, but it should properly be ^^ ^^ piecewise continuously differentiable" or ^^ ^^ piecewise smooth"  to avoid some pathological situations.
}.

\item
When $\vert k \vert \to \infty$ with fixed ${\rm Im}\,k>0$, equation (\ref{1eq3.12}) holds if $f^{(N-1)}$ is continuous and piecewise smooth, and if $\lim_{z\to -\infty}f(z) e^{cz}=0$ for any positive number $c$.

\item
When $\vert k \vert \to \infty$ with ${\rm Im}\,k=0$, equation (\ref{1eq3.12}) holds if $f^{(N-1)}$ is continuous and piecewise smooth, and if $f(-\infty)$ is finite.
\end{enumerate}
This means that if the conditions in (i), (ii) or (iii) hold, then (\ref{1eq3.12}) holds in the region
(i) $\epsilon \leq \arg k \leq \pi-\epsilon$, \ 
(ii) ${\rm Im}\,k \geq \epsilon$, \ 
(iii) ${\rm Im}\,k\geq 0$,\,
respectively, where $\epsilon$ is an arbitrary positive number. (We shall always let $\epsilon$ stand for a positive quantity.)

\section{Expansion of $\bi{S}$}

We wish to derive the $1/k$-expansion of $\log G$ from (\ref{1eq2.15}) by using the formulas given in the previous section.
To do so, we need to study the expansion of $S$ and $\log (1-S)$. 
From (\ref{1eq2.12}) and (\ref{1eq3.1}a), we find
\begin{equation}
\label{1eq4.1}
S_r(x,k)=\lim_{\xi \to -1}\frac{1}{1-\xi^2}\left[\bar R_r(x,-\infty;\xi;k)+\xi \right].
\end{equation}
Therefore, the expansion of $S_r$ is obtained by substituting (\ref{1eq3.3}) into (\ref{1eq4.1}) as
\begin{equation}
\label{1eq4.2}
\fl
S_r(x,k)=\frac{1}{2ik}s_1(x)+\frac{1}{(2ik)^2}s_2(x)+\cdots + \frac{1}{(2ik)^N}s_N(x)
+\sigma^r_N(x,k),
\end{equation}
where, using definition (\ref{1eq3.7}),
\begin{equation}
\label{1eq4.3}
s_n(x)=\tilde c_n(x,-1),
\end{equation}
\begin{equation}
\label{1eq4.4}
\sigma^r_N(x,k)=\lim_{\xi \to -1}\frac{1}{1-\xi^2} \bar \delta_N(x,\xi,k).
\end{equation}
Setting $\xi=-1$ in (\ref{1eq3.10}), we have
\begin{eqnarray}
\label{1eq4.5}
\fl
s_1=-f, \quad
s_2=-f'-f^2, \quad
s_3=-f''-2ff', \quad
s_4=-f'''-2ff''-{f'}^2+2f^2f'+f^4
\nonumber\\
\fl
s_5=-f^{(4)}-2ff'''-2f'f''+4f^2f''+8f(f')^2+8f^3f',
 \quad {\rm etc.}
\end{eqnarray}
Substituting (\ref{1eq3.5}) with (\ref{1eq3.1}) into (\ref{1eq4.4}) gives the expression for $\sigma_N^r$,
\begin{equation}
\label{1eq4.6}
\sigma^r_N(x,k)=\frac{1}{(2ik)^N}\int_{-\infty}^x
\left[\frac{\tau(x,z;k)}{1+R_r(x,z;k)}\right]^2 K_N(z,\eta_l)\,dz,
\end{equation}
where we have defined
\begin{equation}
\label{1eq4.7}
\eta_l\equiv R_l(x,z;k)-\frac{\tau^2(x,z;k)}{1+R_r(x,z;k)}.
\end{equation}

The expressions for $S_l$ corresponding to (\ref{1eq4.2}), (\ref{1eq4.6}) and (\ref{1eq4.7})  can be derived in the same way, using the analogues of the formulas of section~3 for $R_l$. The result is
\begin{equation}
\label{1eq4.8}
\fl
S_l(x,k)=\frac{1}{(-2ik)}s_1(x)+\frac{1}{(-2ik)^2}s_2(x)+\cdots 
 + \frac{1}{(-2ik)^N}s_N(x)
+\sigma^l_N(x,k),
\end{equation}
\begin{equation}
\label{1eq4.9}
\sigma^l_N(x,k)=\frac{-1}{(-2ik)^N}\int_x^\infty
\left[\frac{\tau(z,x;k)}{1+R_l(z,x;k)}\right]^2 K_N(z,\eta_r)\,dz,
\end{equation}
\begin{equation}
\label{1eq4.10}
\eta_r\equiv R_r(z,x;k)-\frac{\tau^2(z,x;k)}{1+R_l(z,x;k)}.
\end{equation}
The coefficients $s_n$ in (\ref{1eq4.8}) are the same ones as in (\ref{1eq4.2}).
The expansion of $S(x,k)$ (equation (\ref{1eq2.13})) is obtained by adding (\ref{1eq4.2}) and (\ref{1eq4.8}):
\begin{equation}
\label{1eq4.11}
\fl
S(x,k)=2\left[\frac{1}{(2ik)^2} s_2(x)+\frac{1}{(2ik)^4} s_4(x)
+\cdots +\frac{1}{(2ik)^{N'}} s_{N'}(x) \right]+\sigma_N(x,k),
\end{equation}
\begin{equation}
\label{1eq4.12}
\sigma_N(x,k) \equiv \sigma_N^r(x,k)+\sigma_N^l(x,k),
\end{equation}
where
\begin{equation}
\label{1eq4.13}
N'\equiv
\cases{
N & ($N$ even) \\
N-1 & ($N$ odd)
}.
\end{equation}
Note that
$\sigma_{N+1}=\sigma_N$ for any even number $N$.

Obviously, a sufficient condition for
\begin{equation}
\label{1eq4.14}
\lim_{\vert k \vert\to \infty} k^N \sigma_N(x,k)=0
\end{equation}
is that the following two equations hold:
\refstepcounter{equation}
\label{1eq4.15}
\addtocounter{equation}{-1}
\begin{equation}
\lim_{\vert k \vert\to \infty} k^N \sigma_N^r(x,k)=0,
\qquad
\lim_{\vert k \vert\to \infty} k^N \sigma_N^l(x,k)=0.
\end{equation}
From (\ref{1eq4.4}) we can see that the first equation of (\ref{1eq4.15}) holds if (\ref{1eq3.12}) holds.
(The limit $\xi \to -1$ does not interfere with the limit $\vert k \vert \to \infty$. We can also show this directly by substituting the expansion of $R_r$ into the first equation of (\ref{1eq2.12}).) Therefore, the first equation of (\ref{1eq4.15}) holds under the conditions  given in section~3. The conditions for the second equation are obvious analogues. 
Hence it is apparent that (\ref{1eq4.14}) holds in the sector $\epsilon \leq \arg k \leq \pi-\epsilon$ if $f^{(N-1)}$ is continuous and piecewise continuously differentiable.
If, in addition, both $f(+\infty)$ and $f(-\infty)$ are finite, then (\ref{1eq4.14}) holds for 
$0 \leq \arg k \leq \pi$.

\section{Expansion of $\mathbf{log(1-\bi{S})}$}

From (\ref{1eq4.11}) we can derive the expansion of $\log(1-S)$ as
\begin{eqnarray}
\label{1eq5.1}
\fl
-\frac{1}{2}\log\left[1-S(x,k)\right]=
\frac{1}{(2ik)^2}\alpha_2(x)+\frac{1}{(2ik)^4}\alpha_4(x)+\cdots
+\frac{1}{(2ik)^{N'}}\alpha_{N'}(x)+\rho_N(x,k),
\nonumber \\
\fl 
\end{eqnarray}
where the coefficients $\alpha_2, \alpha_4, \alpha_6,\ldots$ are expressed in terms of $\{s_n\}$ as
\begin{equation}
\label{1eq5.2}
\alpha_{2i}=\frac{1}{2}\sum_{m=1}^\infty\frac{2^m}{m}
\sum_{\{j_1,\ldots j_m\} \atop \Sigma \,j_n=i}
s_{2j_1}s_{2j_2}\cdots s_{2j_m}.
\end{equation}
(The second sum in (\ref{1eq5.2}) is over $j_n=1,2,3,\ldots$ for each $n$ $(1 \leq n \leq m)$ with the constraint $\sum_{n=1}^m j_n=i$.)
The remainder term $\rho_N$ of (\ref{1eq5.1}) can be written as
\begin{equation}
\label{1eq5.3}
\rho_N=
\frac{1}{2}\sum_{m=1}^\infty \frac{1}{m}
\left(\frac{\sigma_N}{1-A_N}\right)^m
+
\left[
-\frac{1}{2}\log\left(1-A_N\right)-\sum_{n=1}^{N'/2} \frac{\alpha_{2n}}{(2ik)^{2n}}
\right],
\end{equation}
where $A_N\equiv 2\sum_{n=1}^{N'/2} s_{2n}/(2ik)^{2n}$,
so that $S=A_N+\sigma_N$ (see (\ref{1eq4.11})). 
 The quantity in the brackets on the right-hand side of (\ref{1eq5.3}) is $O(1/\vert k \vert^{N+1})$ as $\vert k \vert \to \infty$.
Hence
\begin{equation}
\label{2eq5.4}
\rho_N =\frac{1}{2} \sigma_N[1+o(1)]+O(1/\vert k \vert^{N+1})
\qquad (\vert k \vert \to \infty)
\end{equation}
provided that $\sigma_N$ vanishes as $\vert k\vert \to \infty$. 
Note also that
\begin{equation}
\label{2eq5.5}
\rho_N=\rho_{N+1} \quad (N \ \hbox{even}),
\qquad
\rho_N=\rho_{N+1}+\frac{\alpha_{N+1}}{(2ik)^{N+1}} \quad (N \ \hbox{odd}).
\end{equation}

The coefficients $\alpha_2, \alpha_4, \ldots$ of (\ref{1eq5.1}), which are given by (\ref{1eq5.2}),  can be expressed in a more compact form. From equations (\ref{1eqa.3}) of appendix~A, we obtain
\begin{equation}
\label{1eq5.5}
-\frac{1}{2}\frac{\partial}{\partial x}\log[1-S(x,k)]
= f(x) +ik[S_r(x,k)-S_l(x,k)].
\end{equation}
Substituting (\ref{1eq4.2}) and (\ref{1eq4.8}) into the right-hand side of (\ref{1eq5.5}) yields
\begin{equation}
\label{1eq5.6}
\fl
-\frac{1}{2}\frac{\partial}{\partial x}\log[1-S(x,k)]
=\frac{1}{(2ik)^2}s_3(x)+\frac{1}{(2ik)^4} s_5(x)+\frac{1}{(2ik)^6} s_7(x)+\cdots,
\end{equation}
where we have used $s_1=-f$.
Comparing (\ref{1eq5.1}) with (\ref{1eq5.6}) we find that $(d/dx)\alpha_n(x)=s_{n+1}(x)$ for any even number $n$.
Integrating both sides of this equation gives
\begin{equation}
\label{1eq5.8}
\alpha_n(x)=\int^x s_{n+1}(z)\,dz.
\end{equation}
The integral on the right-hand side is uniquely determined in such a way that $\alpha_n$ includes no additional integration constant when expressed in terms of $f$ and its derivatives. 
For example, $
s_3=-\left(f^2+f'\right)'
$, 
$
s_5=\left(2f^4+4f^2f'-2ff''+f^{(3)}\right)'
$, 
as can be seen from (\ref{1eq4.5}).
In this way, $s_m$ is a total derivative for odd $m\geq 3$.
Hence we have
\begin{equation}
\label{1eq5.10}
\alpha_2=-f^2-f', \qquad
\alpha_4=2f^4+4f^2f'-2ff''+f^{(3)}, \quad {\rm etc.}
\end{equation}
If $f(-\infty)=0$ or $f(+\infty)=0$, we can write the right-hand side of (\ref{1eq5.8}) as 
$\int^x_{-\infty} s_{n+1}(z)\,dz$ or $-\int^{\infty}_x s_{n+1}(z)\,dz$. 
But (\ref{1eq5.8}) holds even if neither $f(-\infty)=0$ nor $f(+\infty)=0$.
Equations (\ref{1eq5.1}) and (\ref{1eq5.5}) also give the relation between the remainder terms
\begin{equation}
\label{1eq5.11}
\frac{\partial}{\partial x}\rho_N (x,k)
=ik \left[\sigma_{N+1}^r(x,k)-\sigma_{N+1}^l(x,k)\right].
\end{equation}

\section{The $\mathbf{1/\bi{k}}$-expansion of $\mathbf{log\,\bi{G}}$}

Substituting (\ref{1eq4.11}), (\ref{1eq5.1}) and (\ref{1eq5.8}) into (\ref{1eq2.15}), we obtain, for any integer $N\geq 0$, 
\begin{eqnarray}
\label{1eq6.1}
\fl
\log G(x,y;k)&=ik(x-y)+\frac{1}{2ik}a_1(x,y)+\frac{1}{(2ik)^2}a_2(x,y)+\frac{1}{(2ik)^3}a_3(x,y)
\nonumber \\
\fl
&\qquad +\cdots+\frac{1}{(2ik)^N}a_N(x,y)+\Delta_N(x,y;k),
\end{eqnarray}
where
\refstepcounter{equation}
\label{1eq6.2}
\addtocounter{equation}{-1}
\numparts
\begin{eqnarray}
a_n(x,y)&=-\int_y^xs_{n+1}(z)\,dz \qquad &(n \ \hbox{odd}),
\\
a_n(x,y)&=
\int^x s_{n+1}(z)\,dz+\int^y s_{n+1}(z)\,dz \qquad &(n \ \hbox{even}),
\end{eqnarray}
\endnumparts
\begin{equation}
\label{1eq6.3}
\Delta_N(x,y;k)= -ik \int_y^x\sigma_{N+1}(z,k)\,dz+\rho_N(x,k)+\rho_N(y,k).
\end{equation}
We can also write the right-hand side of (\ref{1eq6.2}b) as $\alpha_n(x)+\alpha_n(y)$ with  $\alpha_n$ given by (\ref{1eq5.2}).

For (\ref{1eq6.1}) to be meaningful as a high-energy expansion, $\Delta_N$ must satisfy
\begin{equation}
\label{1eq6.5}
\lim_{\vert k \vert \to \infty} k^N \Delta_N(x,y;k)=0.
\end{equation}
Let us study the conditions for (\ref{1eq6.5}) using expression (\ref{1eq6.3}).
We first remark that (\ref{2eq5.4}) and (\ref{2eq5.5}) imply
\begin{equation}
\label{2eq6.6}
\rho_N =\frac{1}{2} \sigma_{N+1}[1+o(1)]+O(1/\vert k \vert^{N+1})
\qquad (\vert k \vert \to \infty).
\end{equation}
Now suppose that $\lim_{\vert k \vert \to \infty} k^{N+1} \sigma_{N+1}=0$. 
Then,  
\begin{equation}
\label{1eq6.6}
\lim_{\vert k \vert \to \infty} k^{N+1} \int_y^x\sigma_{N+1}(z,k)\,dz=0, \qquad
\lim_{\vert k \vert \to \infty} k^N \rho_N=0,
\end{equation}
and so (\ref{1eq6.5}) holds. 
(Since $\vert \sigma_{N+1}(z,k) \vert$ is uniformly bounded in the interval $y<z<x$, the limit
$\vert k \vert \to \infty$ and the integral can be interchanged. The second equation follows from (\ref{2eq6.6}).)
We know that $\lim_{\vert k \vert \to \infty} k^{N+1} \sigma_{N+1}=0$ is satisfied under the conditions stated at the end of section~4, with $N$ replaced by $N+1$.
Hence it follows that (\ref{1eq6.5}) holds for $\epsilon \leq \arg k\leq \pi-\epsilon$ if $f^{(N)}$ is continuous and piecewise continuously differentiable. 
It holds for $0 \leq \arg k \leq \pi$ if, in addition, both $f(-\infty)$ and $f(+\infty)$ are finite. 

\section{Validity of (\ref{1eq6.5}) for discontinuous ${\bi f}^{(N)}$}

The conditions for (\ref{1eq6.5}) mentioned at the end of the last section are sufficient conditions, not necessary ones. 
Here we show that $f^{(N)}$ need not be continuous for (\ref{1eq6.5}) to hold.

Suppose that $f(z)$ is $(M-1)$ times continuously differentiable, $f^{(M-1)}(z)$ is continuously differentiable except at $z=0$, and that $f^{(M)}$ has a finite jump at $z=0$: 
\begin{equation}
\label{1eq7.1}
f^{(M)}(z)=C\theta(z)+\cdots.
\end{equation}
Here $\theta(z)$ is the Heaviside step function, and $C$ is a constant. 
Then $f^{(M+1)}$ contains a delta function as $f^{(M+1)}(z)=C\delta(z)+\cdots$. It is easy to show that $K_n$ (equation (\ref{1eq3.9})) has the form $K_n=f^{(n)}+\cdots$, where the remaining terms on the right-hand side do not contain derivatives of order $n$ or higher. (See (\ref{1eq3.11}).) Therefore,
\begin{equation}
\label{1eq7.3}
K_{M+1}(z,\xi)=C\delta(z)+\cdots.
\end{equation}
Substituting this into (\ref{1eq4.6}) and (\ref{1eq4.9}) with $N=M+1$, and then into (\ref{1eq4.12}), we have
\begin{eqnarray}
\fl
\label{1eq7.4}
\sigma_{M+1}(z,k)&=
\frac{C}{(2ik)^{M+1}}
\int_{-\infty}^z
\left[\frac{\tau(z,w)}{1+R_r(z,w)}\right]^2 \delta(w)\,dw
\nonumber \\
\fl
& \qquad
-
\frac{C}{(-2ik)^{M+1}}\int_z^\infty
\left[\frac{\tau(w,z)}{1+R_l(w,z)}\right]^2 \delta(w)\,dw +\cdots
\nonumber \\
\fl
&=\frac{C}{(2ik)^{M+1}}
\left\{
\left[\frac{\tau(z,0)}{1+R_r(z,0)}\right]^2 \theta(z)
+(-1)^M\left[\frac{\tau(0,z)}{1+R_l(0,z)}\right]^2 \theta(-z)
\right\}+\cdots.
\nonumber \\
\fl
\end{eqnarray}
The quantity in the curly brackets in the last line of (\ref{1eq7.4}) does not vanish as $\vert k \vert \to \infty$ when $k$ is real. (See (\ref{1eqa.2}).) But its integral does vanish in this limit even when ${\rm Im}\,k=0$. Indeed, using (\ref{1eqa.5}), (\ref{1eqa.1}) and (\ref{1eqa.2}) of appendix~A we find
\begin{eqnarray}
\label{2eq7.5}
\fl
&\int_y^x
\left\{
\left[\frac{\tau(z,0)}{1+R_r(z,0)}\right]^2 \theta(z)
+(-1)^M\left[\frac{\tau(0,z)}{1+R_l(0,z)}\right]^2 \theta(-z)
\right\}
dz
\nonumber \\
\fl
&\qquad=\frac{1}{2ik}
\biggl[
\left(e^{2ikx}-e^{2ik\max(y,0)}\right)\theta(x)
+(-1)^M\left(e^{-2iky}-e^{-2ik\min(x,0)}\right)\theta(-y)
\biggr]
\nonumber \\
\fl
& \qquad \qquad
+o(1/\vert k \vert)\qquad (\vert k \vert \to \infty, \ 0 \leq \arg k \leq \pi). 
\end{eqnarray}
Assuming that the part represented by ^^ ^^ $\cdots$" in (\ref{1eq7.4}) is $o(1/\vert k\vert^{M+1})$, we have
\begin{equation}
\label{1eq7.5}
\sigma_{M+1}(z,k)=O(1/\vert k \vert^{M+1}),
\qquad
\int_y^x \sigma_{M+1}(z,k)\,dz=o(1/\vert k \vert^{M+1}),
\end{equation}
and, from (\ref{2eq6.6}),
\begin{equation}
\label{1eq7.6}
\rho_M(z,k)=O(1/\vert k \vert^{M+1}).
\end{equation}
From (\ref{1eq6.3}), (\ref{1eq7.5}) and (\ref{1eq7.6}) it follows that $\Delta_M=o(1/\vert k \vert^M)$, so (\ref{1eq6.5}) holds for $N=M$ in spite of the discontinuity of $f^{(M)}$. 

The same argument holds when $f^{(M)}$ has two or more finite jumps.
In summary, the following conclusion can be drawn:

\medskip
\noindent
\begin{enumerate}
\item
If $f$ is $(N-1)$ times continuously differentiable and $f^{(N-1)}$ is  piecewise continuously differentiable, and if $f^{(N)}$ is piecewise continuously differentiable, then (\ref{1eq6.5}) holds in the angular region $\epsilon \leq \arg k \leq \pi-\epsilon$ with arbitrary $\epsilon>0$.  (Here $f^{(N)}$ may possibly have a finite number of finite jumps.)
\item
If the conditions of (i) are satisfied, and if $\lim_{z\to +\infty} f(z) e^{-cz}=\lim_{z\to -\infty}f(z) e^{cz}=0$ for any $c>0$, then (\ref{1eq6.5}) holds in the half plane ${\rm Im}\,k\geq \epsilon$ with arbitrary $\epsilon>0$. 
\item
If the conditions of (i) are satisfied, and if both $f(+\infty)$ and $f(-\infty)$ are finite, then (\ref{1eq6.5}) holds in the upper half plane ${\rm Im}\,k\geq 0$ including the real axis.
\end{enumerate}
(As noted below equation (\ref{1eq3.12}), we are assuming that $f(z)$ and its derivatives are monotone for sufficiently large $\vert z \vert$, at both $z \to +\infty$ and $z \to -\infty$.)

The conditions of (i) are still not necessary conditions. Actually, (\ref{1eq6.5}) may hold even when $f$ is not $(N-1)$ times differentiable, as will be shown in the next section.

\section{Effects of the jump of ${\bi f}^{(M)}$ at higher orders}

Here we show that (\ref{1eq6.5}) may hold even for $N\geq M+1$ when $f^{(M)}$ has a discontinuity as in (\ref{1eq7.1}).  
To this end, let us study $\Delta_{M+1}, \Delta_{M+2}, \cdots$ for the case of (\ref{1eq7.1}). Since $K_{M+2}=f^{(M+2)}+\cdots$, we have
\begin{equation}
\label{1eq8.1}
K_{M+2}(z,\xi)=C\delta'(z)+\cdots,
\end{equation}
where $\delta'$ is the derivative of the delta function.
As before, we substitute (\ref{1eq8.1}) into (\ref{1eq4.6}) and (\ref{1eq4.9}), and then into   (\ref{1eq4.12}). After integrating by parts, we obtain
\begin{eqnarray}
\label{2eq8.2}
\fl
\sigma_{M+2}(z,k)&=
\frac{C}{(2ik)^{M+2}}
\left\{
\delta (z)
-\left.\frac{\partial}{\partial w}\left[\frac{\tau(z,w)}{1+R_r(z,w)}\right]^2 \right\vert_{w=0}\theta(z)
\right\}
\nonumber \\
\fl
& \qquad
+\frac{C}{(-2ik)^{M+2}}
\left\{
\delta (z)
+\left.\frac{\partial}{\partial w}\left[\frac{\tau(w,z)}{1+R_l(w,z)}\right]^2 \right\vert_{w=0}\theta(-z)
\right\}
+\cdots.
\end{eqnarray}
Using (\ref{1eqa.4}) of appendix~A, we can write (\ref{2eq8.2}) as 
\begin{eqnarray}
\label{1eq8.2}
\fl
\sigma_{M+2}(z,k)&=\frac{C}{(2ik)^{M+1}}
\left\{
\left[\frac{\tau(z,0)}{1+R_r(z,0)}\right]^2 \theta(z)
+(-1)^M\left[\frac{\tau(0,z)}{1+R_l(0,z)}\right]^2 \theta(-z)
\right\}
\nonumber \\
\fl
& \qquad +\frac{2\,C}{(2ik)^{M+2}}
A_M\delta(z)
+\cdots,
\end{eqnarray}
where
\begin{equation}
A_M\equiv
\cases{
1 & ($M$ even) \\
0 & ($M$ odd)
}.
\end{equation}
In (8.3), we have disregarded the terms of order $1/k^{M+2}$ which have the form $Q(z,k)/(2{\rm i}k)^{M+2}$ with $Q(z,k)$ such that $\int_y^x Q(z,k) d z$ vanishes as $\vert k \vert \to \infty$. Such terms are irrelevant to the following discussion.
In addition to $C\delta'(z)$, the right-hand side of (8.1) contains another singular term proportional to $f(z)\delta(z)$, but the contribution from this term can be disregarded for the same reason. 
We assume that the contribution to (8.3) from non-singular terms of $K_{M+2}$ is $o(1/\vert k \vert^{M+2})$ as $\vert k \vert \to \infty$.

Now we consider the two cases, $y<0<x$ and $0<y<x$.

\bigskip
\noindent
(a) \ $y<0<x$

\nopagebreak
\smallskip
\noindent
The integral of (\ref{1eq8.2}) can be calculated by using (\ref{2eq7.5}).
For the case $y<0<x$ we have
\begin{equation}
\label{1eq8.3}
\int_y^x\sigma_{M+2}(z,k)\,dz=
\frac{C}{(2ik)^{M+2}}
\left[
e^{2ikx}+(-1)^M e^{-2iky}
\right]
+o(1/\vert k\vert^{M+2}).
\end{equation}
(Note that the contribution from the delta function in (\ref{1eq8.2}) is canceled.) 
From (\ref{2eq5.4}), (\ref{1eq7.4}) and (\ref{1eqa.2}) (or from (\ref{2eq6.6}), (\ref{1eq8.2}) and (\ref{1eqa.2})) we obtain
\refstepcounter{equation}
\label{1eq8.4}
\addtocounter{equation}{-1}
\numparts
\begin{equation}
\rho_{M+1}(x,k)=\frac{C}{2(2ik)^{M+1}}\,e^{2ikx}
+o(1/\vert k\vert^{M+1}),
\end{equation}
\begin{equation}
\rho_{M+1}(y,k)=(-1)^M \frac{C}{2(2ik)^{M+1}}\,e^{-2iky}
+o(1/\vert k\vert^{M+1}).
\end{equation}
\endnumparts
On substituting (\ref{1eq8.3}), (\ref{1eq8.4}a) and (\ref{1eq8.4}b) into (\ref{1eq6.3}) with $N=M+1$, the terms of order $1/k^{M+1}$ cancel out, and we have
\begin{equation}
\label{1eq8.5}
\Delta_{M+1}(x,y;k)=o(1/\vert k\vert^{M+1})
\qquad (\vert k \vert \to \infty).
\end{equation}
Namely, (\ref{1eq6.5}) holds for $N=M+1$. 

The effect of the discontinuity of $f^{(M)}$ on $\Delta_{M+2},\Delta_{M+3}$ etc can be studied in the same way.
Singular terms involving derivatives of the delta function such as $C\delta'', C\delta''',\ldots$ appear in $K_{M+3}, K_{M+4}, \ldots$, respectively, but they cause no problems. 
Singular contributions coming from these terms cancel out in the expressions for $\Delta_{M+2},\Delta_{M+3}$ etc, as we have seen above for $\Delta_{M+1}$. 
(The explanation is omitted here, but the cancellation between $ik\int\sigma_{N+1}\,dz$ and $\rho_N$ is guaranteed by relation (\ref{1eq5.11}).)
As a result, the derivatives of the delta function do not produce a term of order $1/k^N$ (or lower) in $\Delta_N$. 
In (\ref{1eq6.2}) and (\ref{1eq6.3}), the delta function and its derivatives appear only within an integral, and these expressions are well defined\footnote{
Needless to say, 
$\int_a^b g(z) \delta^{(m)}(z) dz=(-1)^m g^{(m)}(0)$ for $a<0<b$. 
In (\ref{1eq6.2}) and (\ref{1eq6.3}), the function multiplying $\delta^{(m)}$ is always $m$ times continuously differentiable if $n\leq 2M+1$ (for (\ref{1eq6.2})) or $N\leq 2M$ (for (\ref{1eq6.3})).
An expression like $\int_a^b \theta(z)g(z)\delta(z)dz=\theta(0)g(0)$ may appear in $\Delta_{2M+1}$, where $\theta(0)=1/2$.
}
 as long as $n, N\leq 2M+1$.

However, (\ref{1eq6.3}) is not well defined for $N=2M+2$ since the square of the delta function appears in it.
We can easily see that $K_{2M+3}$ contains a term proportional to $(f^{(M+1)})^2=C^2\delta^2$.
In addition, terms proportional to $f^{(M+1+i)}f^{(M+1-i)}$ ($i=1,2,\ldots$) are contained in $K_{2M+3}$. These terms are ill defined, too, since they produce the square of the delta function by integration by parts. 
To study $\Delta_{2M+2}$, we need to go back one step, as we cannot directly use (\ref{1eq6.3}). 
By induction, it can be shown that
\begin{equation}
\label{2eq8.9}
K_{2M+2}=f^{(2M+2)}-4\xi\sum_{i=1}^{M+1}f^{(M+i)}f^{(M+1-i)}+\cdots.
\end{equation}
The terms explicitly shown in (\ref{2eq8.9}) are the ones that give rise to the ill-defined  terms in $K_{2M+3}$. We substitute (\ref{2eq8.9}) into the expression for $\Delta_{2M+1}$. Then we can study $\Delta_{2M+2}$ by using this expression together with the relation $\Delta_{2M+2}=\Delta_{2M+1}-a_{2M+2}/(2ik)^{2M+2}$. After some calculation\footnote{
It turns out that only the first term on the right-hand side of (\ref{2eq8.9}) is relevant.
Anomalous contribution comes from where $\theta(0)$ appears in formal calculation whereas it should really be $\theta(0+\epsilon)$ or $\theta(0-\epsilon)$.
For example, 
 $ik \!\int_0^x\theta(z)g(z)e^{ik z}dz=-g(0)+g(x)e^{ikx}+o(1)$ 
 $(\vert k \vert \to \infty)$, where 
 the first term on the right-hand side is not $-\theta(0)g(0)=-g(0)/2$ but $-\theta(0+\epsilon)g(0)$.
In deriving (\ref{2eq8.10}), we also use the $1/k$-expansion of $R_r$, $R_l$ and $\tau$ for finite intervals. (See the comments below (\ref{1eqa.2}) in appendix~A.)
In the expansion, we need only to keep track of the terms which are linear in $f$ and its derivatives.
}
  we find that $\Delta_{2M+2}$ contains a term of order $1/k^{2M+2}$ as
\begin{equation}
\label{2eq8.10}
\Delta_{2M+2}=\frac{1}{(2ik)^{2M+2}}\frac{(-1)^M}{2}C^2+ \cdots.
\end{equation}
So (\ref{1eq6.5}) does not hold for $N=2M+2$.
We can see that (\ref{2eq8.10}) gives correction to the coefficient of order $2M+2$ as
\begin{eqnarray}
\label{2eq8.11}
\fl
\log G=ik(x-y)+\frac{1}{2ik}a_1+
\cdots+\frac{1}{(2ik)^{2M+2}}
\left[a_{2M+2}+(-1)^M C^2/2\right]+\cdots.
\end{eqnarray}
Thus, the correct coefficient of order $2M+2$ is not $a_{2M+2}$ but $a_{2M+2}+(-1)^M C^2/2$.

In summary, when $f^{(M)}$ has a finite jump of the form of (\ref{1eq7.1}), and when the jump is located between $y$ and $x$, equation (\ref{1eq6.5}) holds for $N\leq 2M+1$ in the sector $\epsilon\leq \arg k\leq \pi-\epsilon$ provided that the part represented by ^^ ^^ $\cdots$" in (\ref{1eq7.1}) is sufficiently differentiable. If, in addition, $f(-\infty)$ and $f(+\infty)$ are both finite, then (\ref{1eq6.5}) holds for $N\leq 2M+1$ even for real $k$.
In any case, (\ref{1eq6.5}) does not hold for $N\geq 2M+2$. 
Although $\log G$ can be asymptotically expanded in powers of $1/k$ even beyond the term of order $1/k^{2M+2}$, the coefficients $a_n$ given by (\ref{1eq6.2}) are not correct for $n\geq 2M+2$. In particular, $a_{2M+2}$ is shifted by $(-1)^M C^2/2$ as in (\ref{2eq8.11}). (See examples 5--8 of section~11.)

\bigskip
\noindent
(b) \ $0<y<x$

\nopagebreak
\smallskip
\noindent
Similarly, for $0<y<x$ we have, by using (\ref{2eq7.5}),
\begin{equation}
\label{1eq8.6}
\int_y^x\sigma_{M+2}(z,k)\,dz=
\frac{C}{(2ik)^{M+2}}
\left(e^{2ikx}-e^{2iky}\right)
+o(1/\vert k\vert^{M+2}),
\end{equation}
\begin{equation}
\label{1eq8.7}
\rho_{M+1}(y,k)=\frac{C}{2(2ik)^{M+1}}\,e^{2iky}+o(1/\vert k\vert^{M+1}),
\end{equation}
while the expression for $\rho_{M+1}(x,k)$ is the same as (\ref{1eq8.4}a). 
Substituting (\ref{1eq8.6}), (\ref{1eq8.4}a) and (\ref{1eq8.7}) into (\ref{1eq6.3}), we obtain
\begin{equation}
\label{1eq8.8}
\Delta_{M+1}(x,y;k)=\frac{C}{(2ik)^{M+1}}\,e^{2iky}+o(1/\vert k\vert^{M+1})
\qquad (\vert k \vert \to \infty).
\end{equation}

If the limit $\vert k \vert \to \infty$ is taken with fixed $\arg k$ in the sector $\epsilon \leq \arg k \leq \pi-\epsilon$, then $\Delta_{M+1}=o(1/\vert k\vert^{M+1})$ since $e^{2iky}$ falls off exponentially.
The same can be said for $\Delta_{M+2}$, $\Delta_{M+3}$, and so on; the contribution to $\Delta_N$ coming from the discontinuity of $f^{(M)}$ is exponentially small at large $\vert k \vert$ as long as $\epsilon \leq \arg k \leq \pi-\epsilon$.

When $k$ is real, however, $k^{M+1}\Delta_{M+1}$ does not vanish as $\vert k \vert \to \infty$. So, unlike case~(a), equation (\ref{1eq6.5}) does not hold for $N\geq M+1$ when the limit is taken along the real axis. (See example~6 of section~11.)

\bigskip
The case $y<x<0$ can be treated in exactly the same way as (b).
It is straightforward to extend the arguments of (a) and (b) above to the cases where there are two or more such discontinuities. 

\section{Short-time expansion of the Green function}

The expansion of $G$ is obtained by exponentiating (\ref{1eq6.1}) as
\begin{equation}
\label{1eq9.1}
\fl
G=e^{ik(x-y)}
\left[
1+\frac{1}{2ik}b_1+\frac{1}{(2ik)^2}b_2+\cdots +\frac{1}{(2ik)^N}b_N+\Delta'_N
\right],
\end{equation}
where
\begin{equation}
\label{1eq9.2}
b_n(x,y)=\sum_{m=1}^\infty \frac{1}{m!}\sum_{\Sigma j_i=n}a_{j_1} a_{j_2} \cdots a_{j_m},
\end{equation}
\begin{equation}
\label{1eq9.3}
\Delta_N'\equiv
\exp\left[\sum_{m=1}^N\frac{a_m}{(2ik)^m}+\Delta_N\right]
-\sum_{m=0}^N \frac{b_m}{(2ik)^m}.
\end{equation}
(Here the prime does not denote a derivative.) 
It is obvious that $\Delta'_N=o(1/\vert k\vert^N)$ as $\vert k \vert \to \infty$ 
if $\Delta_N=o(1/\vert k\vert^N)$. 

The time dependent Green function for the Fokker-Planck equation is obtained from $G(x,y;k)$ by the inverse Fourier transformation as
\begin{equation}
\label{1eq9.4}
G_{\rm F}(x,y;t)=-\frac{1}{2\pi}e^{-[V(x)-V(y)]/2}
\int_{-\infty}^\infty \frac{1}{2i\kappa}G(x,y;\kappa)e^{-i \omega t}d\omega,
\end{equation}
where $\kappa$ is defined by
\begin{equation}
\label{1eq9.5}
\kappa^2\equiv i\omega, \qquad {\rm Im}\,\kappa\geq 0.
\end{equation}
(When necessary, the integral in (\ref{1eq9.4}) is to be understood as $\int_{-\infty+i\epsilon}^{\infty +i\epsilon}$ with positive infinitesimal $\epsilon$.)
By substituting (\ref{1eq9.1}) into (\ref{1eq9.4}) and carrying out the integration term by term, we can derive an expansion of $G_{\rm F}$ in powers of $t$. The result is
\begin{eqnarray}
\label{1eq9.6}
\fl
G_{\rm F}(x,y;t)=\frac{e^{-[V(x)-V(y)]/2}}{\sqrt{4 \pi t}}
\exp\left[-\frac{(x-y)^2}{4t}\right]
\left(1+g_1t+g_2 t^2 + g_3 t^3+\cdots+g_N t^N+\tilde\Delta_N\right),
\nonumber \\
\end{eqnarray}
where
\begin{equation}
\label{1eq9.7}
g_n(x,y)=\sum_{m=1}^n
\frac{(-1)^n(2n-m-1)!}{(m-1)!\,(n-m)!}
\frac{b_m(x,y)}{(x-y)^{2n-m}}.
\end{equation}
(See appendix~B for the derivation. The expression for $\tilde \Delta_N$ is shown there.)
When $x \neq y$, the remainder term $\tilde \Delta_N$ satisfies
\begin{equation}
\label{1eq9.8}
\lim_{t\to 0} \frac{\tilde \Delta_N(x,y;t)}{t^N}=0
\end{equation}
if (\ref{1eq6.5}) is satisfied for $\epsilon \leq \arg k \leq \pi-\epsilon$. 
(See appendix~B.) 
For (\ref{1eq9.8}) to hold, it is not necessary that (\ref{1eq6.5}) hold for ${\rm Im}\,k=0$. 
 So this short-time expansion is valid even when $f(\pm \infty)$ are infinite. (Here we are assuming that $t$ is real.)

Note that the right-hand side of (\ref{1eq9.7}) is not infinite at $x=y$ in spite of the appearance of the $1/(x-y)^{2n-m}$. For $x=y$ we have
\begin{equation}
\label{1eq9.11}
g_n(x,x)=\frac{b_{2n}(x,x)}{2^n (2n-1)!!},
\end{equation}
as can be directly calculated. The right-hand side of (\ref{1eq9.7}) approaches (\ref{1eq9.11}) as $y \to x$.

\section{Application to the Schr\"odinger equation}

From (\ref{1eq4.5}), we may note that $s_n$ for $n \geq 2$ can be expressed in terms of the Schr\"odinger potential $V_{\rm S}$ (equation~(\ref{1eq1.5})) and its derivatives
\begin{equation}
\label{1eq10.1}
\fl
s_2=-V_{\rm S}, \qquad 
s_3=-V_{\rm S}', \qquad 
s_4=V_{\rm S}^2-V_{\rm S}'', \qquad
s_5=(2 V_{\rm S}^2-V_{\rm S}'')', \quad {\rm etc}.
\end{equation}
This can be confirmed as follows.  
From (\ref{1eqa.3}c) of appendix~A and (\ref{1eq2.12}), we can show that $S_r$ satisfies the differential equation
\begin{equation}
\label{1eq10.2}
\frac{\partial}{\partial x}S_r(x,k)
=2ikS_r(x,k)[1-S_r(x,k)]+f(x)[1-2S_r(x,k)].
\end{equation}
Substituting (\ref{1eq4.2}) into (\ref{1eq10.2}), we obtain
\begin{equation}
\label{1eq10.3}
s_n'=s_{n+1}-\sum_{j=1}^ns_js_{n+1-j}-2f s_n+f\delta_{n0}.
\end{equation}
Hence it follows that the $s_n$'s satisfy the recursion relation
\begin{equation}
\label{1eq10.4}
s_{n+1}=s_n'+\sum_{j=2}^{n-1} s_j s_{n+1-j} \qquad (n \geq 2),
\end{equation}
where we have used $s_1=-f$. 
We can obtain $s_3, s_4, s_5,\ldots$ from this recursion relation, starting with $s_2=-V_{\rm S}$.
Therefore, any $s_n$ ($n\geq 2$) can indeed be expressed in terms of $V_{\rm S}$ and its derivatives.  
Substituting (\ref{1eq10.1}) into (\ref{1eq6.2}), we have
\begin{eqnarray}
\label{2eq10.5}
\fl
a_1=\int_y^x V_{\rm S}(z)\,dz, \qquad
a_2=-V_{\rm S}(x)-V_{\rm S}(y), \qquad
a_3=-\int_y^x\left[V_{\rm S}^2(z)-V_{\rm S}''(z)\right]dz,
\nonumber \\
\fl
a_4=2V_{\rm S}^2(x)-V_{\rm S}''(x)+2V_{\rm S}^2(y)-V_{\rm S}''(y), \quad {\rm etc}.
\end{eqnarray}
Substituting (\ref{2eq10.5}) into (\ref{1eq6.1}), and returning to definition (\ref{1eq2.7}), we can write 
\begin{eqnarray}
\label{1eq10.5}
\fl
\log G_{\rm S}(x,y;E)=-\log 2i -\log k+ik(x-y)
+\frac{1}{2ik}\int_y^x V_{\rm S}(z) \,dz
\nonumber \\
 \qquad -\frac{1}{(2ik)^2}\left[V_{\rm S}(x)+V_{\rm S}(y)\right]
-\frac{1}{(2ik)^3}\int_y^x\left[V_{\rm S}^2(z)-V_{\rm S}''(z)\right]dz+\cdots.
\end{eqnarray}

The recursion relation (\ref{1eq10.4}) is familiar in soliton theory. 
The quantities $s_n$ obtained from it are, apart from the sign, identical to the conserved charge densities for the KdV equation \cite{gesztesy1,novikov}. It is known that these quantities also appear in the asymptotic expansion for Jost solutions \cite{rybkin1,newell}. 
Our results for the expansion of the Green function are valid even when Jost solutions do not exist.

So far, we have been assuming that the Schr\"odinger equation was derived from the Fokker-Planck equation (\ref{1eq1.3}).
Let us now go in the opposite direction. 
To transform a given Schr\"odinger equation into a Fokker-Planck equation, it is, in general,  necessary to shift the energy level. Let $\psi_0(x)$ be a solution of (\ref{1eq1.4}) with $E=E_0$. 
Then the Schr\"odinger equation (\ref{1eq1.4}) is equivalent to the Fokker-Planck equation
\begin{equation}
\label{1eq10.6}
-\frac{d^2}{dx^2}\phi(x)+2\frac{d}{dx}[f(x)\phi(x)]=p^2\phi(x),
\end{equation}
where
\begin{equation}
\label{1eq10.7}
\fl
\phi(x)=\psi_0(x)\psi(x), \qquad
f(x)=\frac{d}{dx}\log \psi_0(x), \qquad
p\equiv \sqrt{k^2-E_0}, \qquad
k^2=E.
\end{equation}
The relation between $V_{\rm S}$ and $f$ is now
\begin{equation}
\label{1eq10.8}
V_{\rm S}(x)-E_0=f^2(x)+f'(x).
\end{equation}
So (\ref{1eq1.5}) is a special case of (\ref{1eq10.8}) with $E_0=0$.
If we want $f(x)$ to be real and finite for any finite $x$, we need to choose $E_0$ such that $\psi_0(x)> 0$ for any finite $x$. 
We may take $E_0$ to be the ground state energy if $V_{\rm S}$ has a bound state.

Equation (\ref{1eq10.5}) was derived by using (\ref{1eq1.5}), which corresponds to $E_0=0$. However, since the Schr\"odinger equation (\ref{1eq1.4}) is invariant under the replacements $V_{\rm S} \to V_{\rm S}-E_0$ and $k \to p$, expansion (\ref{1eq10.5}) is valid even when $E_0\neq 0$. 
We can check this by an explicit calculation. 
Suppose that $E_0 \neq 0$. 
Then it is obvious that a correct expression for $\log G_{\rm S}$ is obtained by making the replacements $V_{\rm S} \to V_{\rm S}-E_0$ and $k \to p$ in (\ref{1eq10.5}) as
\begin{eqnarray}
\label{1eq10.9}
\fl
\log G_{\rm S}(x,y;E)&=-\log 2i -\log p+ip(x-y)
+\frac{1}{2ip}\int_y^x \left[V_{\rm S}(z)-E_0\right] dz
\nonumber \\
\fl
& \qquad 
-\frac{1}{(2ip)^2}\left[V_{\rm S}(x)+V_{\rm S}(y)-2E_0\right]
+\cdots.
\end{eqnarray}
If we rearrange (\ref{1eq10.9}) into an expansion in powers of $1/k$ by substituting
\begin{equation*}
\label{1eq10.10}
\fl
\log p=\log k-\frac{E_0}{2k^2}-\frac{E_0^2}{4k^4}+\cdots, \quad
p=k-\frac{E_0}{2k}-\frac{E_0^2}{8k^3}+\cdots,
\quad
\frac{1}{p}=\frac{1}{k}+\frac{E_0}{2k^3}+\cdots, 
\end{equation*}
and so on, then it becomes (\ref{1eq10.5}). Thus, (\ref{1eq10.5}) holds for $E_0 \neq 0$ as well. Hence we know that (\ref{1eq6.1}), too, is valid for $E_0 \neq 0$.
(See example~8 of the next section. Note that $G$ is not $2ipG_{\rm S}$ but $2ikG_{\rm S}$.) 
Unlike the coefficients $a_n$, the remainder term $\Delta_N$ of (\ref{1eq6.1}) is not expressed solely in terms of $V_{\rm S}$. 
But the conditions for the validity of (\ref{1eq6.5}) remain unchanged when $E_0 \neq 0$. We can understand this by writing (\ref{1eq10.9}) with the remainder term, and then turning  it into an expansion in powers of $1/k$ as above. 

Is is not difficult to interpret the results of sections 7 and 8 in the language of the Schr\"odinger equation. 
We see from the second equation of (\ref{1eq10.7}) that the conditions on $f$ can be interpreted as the conditions on $\psi_0$. 
For example, $f(\infty)$ takes a nonzero finite value when the ground state wave function decays like $e^{-cx}$ as $x \to \infty$. 
The differentiability conditions on $f$ can be directly related to the differentiability of $V_{\rm S}$. Namely, $f$ is $n$~times differentiable if $V_{\rm S}$ is $(n-1)$~times differentiable. 
\section{Examples}

Let us consider some simple potentials for which the exact Green function can be obtained, and compare the exact expression with our results for the expansion
\begin{equation}
\label{1eq11.1}
\log G(x,y;k)= ik(x-y) +  \frac{a_1}{2ik} + \frac{a_2}{(2ik)^2}  + \frac{a_3}{(2ik)^3}  + \cdots.
\end{equation}
The coefficients $a_n$ are obtained from (\ref{1eq6.2}), (\ref{1eq4.3}) and (\ref{1eq3.8}). Explicit forms of the quantities $s_n$ (equation (\ref{1eq4.3})) are shown in (\ref{1eq4.5}) for $n \leq 5$ . 
(Alternatively, we may use expressions (\ref{2eq10.5}) in terms of $V_{\rm S}$.) We omit the derivation of the exact expressions.

\bigskip
\noindent
{\bf Example 1.}\quad
$V(z)=2 z, \quad f(z)=-1, \quad V_{\rm S}(z)=1.$

\nopagebreak
\smallskip
\noindent
As the simplest example, let us first consider a linear potential. 
Substituting $f=-1$ into (\ref{1eq4.5}), and then into (\ref{1eq6.2}), we obtain the coefficients $a_n$ as
\begin{equation}
\label{1eq11.2}
\fl
a_1=x-y, \qquad a_2=-2, \qquad a_3=-x+y, \qquad a_4=4,  \quad {\rm etc}.
\end{equation}
The exact Green function for this potential is
\begin{equation}
\label{1eq11.3}
G(x,y;k)=\frac{k}{\sqrt{k^2-1}} \exp\left[{i\sqrt{k^2-1}(x-y)}\right].
\end{equation}
It is easy to see that (\ref{1eq11.1}) with (\ref{1eq11.2}) is the correct expansion of the logarithm of (\ref{1eq11.3}) for ${\rm Im}\,k\geq 0$. Since the exact $\log G(k)-ik(x-y)$ does not have any singularities in $\vert k \vert >1$, the right-hand side of (\ref{1eq11.1}) is a convergent infinite series for $\vert k \vert >1$. 

The short-time expansion (\ref{1eq9.6}) with $g_n$ calculated from (\ref{1eq9.7}) and (\ref{1eq9.2}) reads
\begin{equation}
\label{1eq11.4}
\fl
G_{\rm F}(x,y;t)= e^{-(x-y)} \frac{1}{\sqrt{4 \pi t}} \exp\left[-\frac{(x-y)^2}{4 t}\right] 
\left( 1 - t + \frac{t^2}{2}  -\frac{t^3}{6}  + \cdots\right).
\end{equation}
The series in parentheses on the right-hand side is $\sum_{n=0}^\infty (-t)^n/n!=e^{-t}$, so this expansion is convergent for any~$t$.

\bigskip
\noindent
{\bf Example 2.}\quad $V(z)=z^2, \quad f(z)=-z, \quad V_{\rm S}(z)=z^2-1.$

\nopagebreak
\smallskip
\noindent
Next, we consider a parabolic potential. 
This example satisfies conditions (i) and (ii) of section~7 for any $N$ but not (iii).
From (\ref{1eq6.2}) and (\ref{1eq4.5}), we obtain
\begin{eqnarray}
\label{1eq11.6}
\fl
&a_1=\frac{x^3-y^3}{3}-(x-y) , \qquad
a_2=2-(x^2+y^2), 
\nonumber \\
\fl
&a_3=-\frac{x^5-y^5}{5}+\frac{2 (x^3-y^3)}{3} + x - y , \qquad
a_4=2 (x^4+y^4)- 4 (x^2 + y^2), \quad {\rm etc}.
\end{eqnarray}
The exact Green function for this potential can be expressed as
\begin{equation}
\label{1eq11.7}
G(x,y;k)=\frac{-ik}{2 \Gamma\left(-\frac{k^2}{4}\right) 
\Gamma\left(\frac{1}{2}-\frac{k^2}{4}\right)} \psi_+(x,k) \psi_-(y,k),
\end{equation}
where
\begin{eqnarray}
\fl
\psi_\pm(x,k)\equiv e^{-x^2/2} 
\left[
\Gamma\left(\textstyle{-\frac{k^2}{4}}\right) 
F\left(\textstyle{-\frac{k^2}{4},\frac{1}{2};x^2}\right)
\mp
2 x 
\Gamma\left(\textstyle{\frac{1}{2}-\frac{k^2}{4}}\right) 
F\left(\textstyle{\frac{1}{2}-\frac{k^2}{4},\frac{3}{2};x^2}\right)
\right].
\end{eqnarray}
Here $\Gamma$ is the gamma function, and $F$ is the confluent hypergeometric function defined by $F(\alpha,\gamma;z)
=\sum_{n=0}^\infty
\frac{\alpha(\alpha+1)\cdots(\alpha+n-1)}{\gamma(\gamma+1)\cdots(\gamma+n-1)}
\frac{1}{n!}z^n
$.
We can check that (\ref{1eq11.1}) with (\ref{1eq11.6}) is the correct asymptotic expansion of the logarithm of (\ref{1eq11.7}) when $\arg k$ is fixed in $0<\arg k<\pi$ (figure~1(a)). Now (\ref{1eq11.1}) is divergent as an infinite series.  
\begin{figure}
\hspace{1cm}
\includegraphics[scale=0.7]{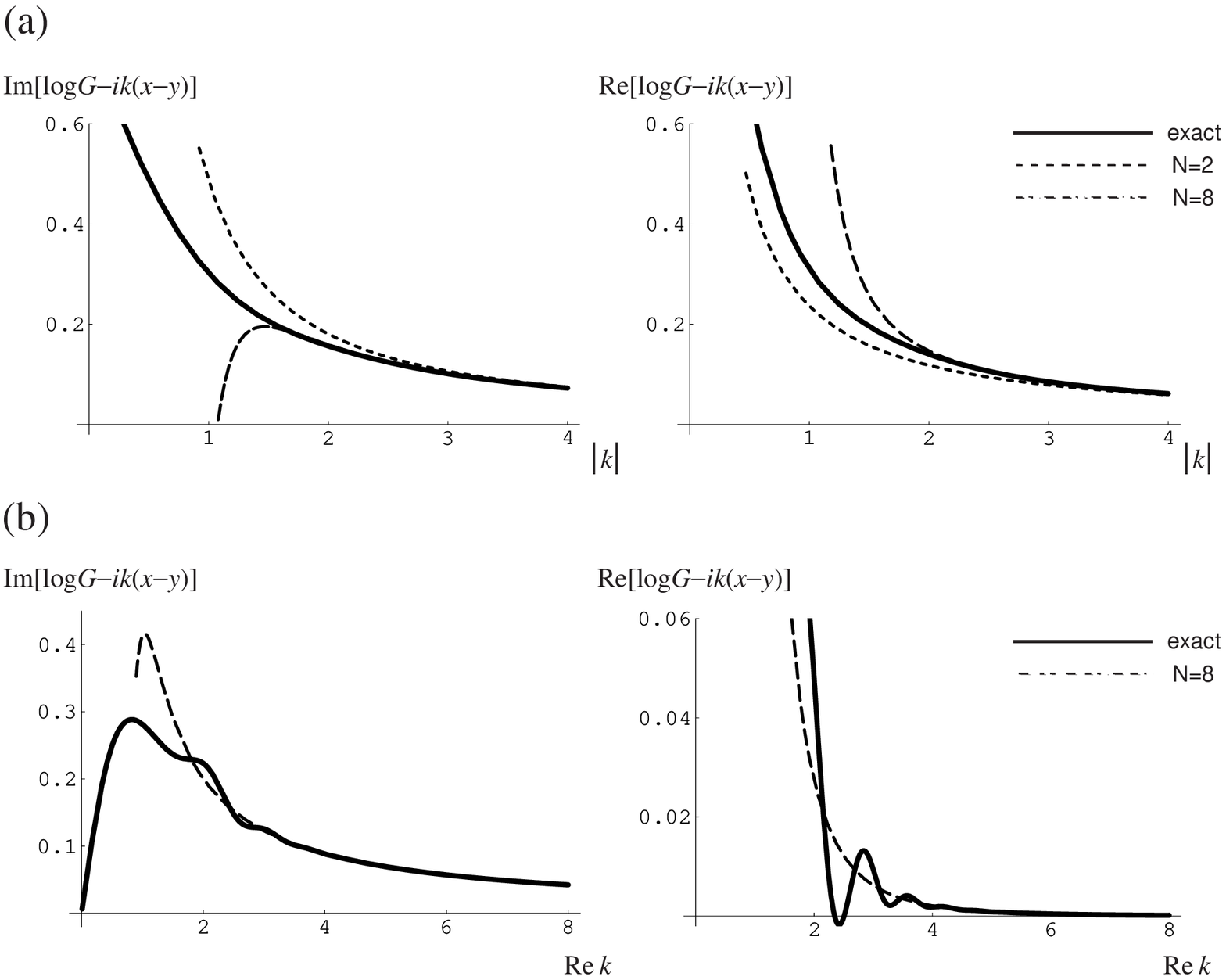}
\caption{
The imaginary and real parts of $\log G(x,y;k)-ik(x-y)$ for the potential $V(z)=z^2$ (example~2), with $x=1$ and $y=0$, 
(a) plotted as functions of $\vert k \vert$, with $\arg k$ fixed at $\pi/4$;
(b) plotted as functions of ${\rm Re}\,k$, with ${\rm Im}\,k$ fixed at $0.75$.
Solid lines: the exact values.
Dashed lines: the expansion (\ref{1eq11.1}) to order $1/k^N$ ($N=2$ and $8$).
}
\end{figure}
This asymptotic expansion is also valid when ${\rm Im}\,k>0$ is kept fixed as $\vert k \vert \to \infty$ (figure~1(b)).
When ${\rm Im}\,k=0$, however, (\ref{1eq11.1}) is not valid since $\log G(k)$ oscillates and does not vanish as $\vert k \vert \to \infty$. 

The short-time expansion (\ref{1eq9.6}) now reads
\begin{equation}
\label{1eq11.9}
\fl
G_{\rm F}(x,y;t)=\frac{1}{\sqrt{4\pi t}}\exp \left[\frac{-x^2+y^2}{2}-\frac{(x-y)^2}{4 t}\right] 
\left(1+g_1t+g_2 t^2 + g_3 t^3 +\cdots \right),
\end{equation}
where the $g_n$'s are obtained from (\ref{1eq9.7}), (\ref{1eq9.2}) and (\ref{1eq11.6}) as
\begin{eqnarray}
\label{1eq11.10}
\fl
&g_1=1-\frac{1}{3}(x^2+xy+y^2), \qquad
g_2=\frac{1}{6}-\frac{1}{3}(x^2+xy+y^2)+\frac{1}{18}(x^2+xy+y^2)^2, 
\nonumber \\
\fl
&g_3=-\frac{1}{6} + \frac{1}{15} xy + \frac{1}{30}(x^2+xy+y^2)+ \frac{1}{18}(x^2+xy+y^2)^2 
-\frac{1}{162}(x^2+xy+y^2)^3.
\end{eqnarray}
The exact time-dependent Green function has the well-known form
\begin{equation}
\label{1eq11.11}
\fl
G_{\rm F}(x,y;t)=\left(\frac{1}{\pi [1-\exp(-4t)]}\right)^{1/2}
\exp
\left(
-\frac{[x-y\exp(-2t)]^2}{1-\exp(-4t)}
\right).
\end{equation}
The expansion of (\ref{1eq11.11}) indeed has the form of (\ref{1eq11.9}) with (\ref{1eq11.10}). 
This $G_{\rm F}(t)$ has singularities in the complex $t$ plane where $\exp(-4t)=1$. The nearest singularities to the origin are at $t=\pm \pi i/2$. 
So, the infinite series (\ref{1eq11.9}) is convergent for $t<\pi/2$. This is a typical case where the short-time expansion is convergent for small~$t$ although the high-energy expansion, from which it was derived, is divergent. 

\bigskip
\noindent
{\bf Example 3.}\quad $V(z)=e^z, \quad f(z)=-e^z/2, 
\quad V_{\rm S}(z)=(e^{2z}/4)-(e^z/2).$

\nopagebreak
\smallskip
\noindent
This exponential potential satisfies conditions (i) of section~7 for any $N$ but not (ii) or (iii). 
For this potential we have
\begin{eqnarray}
\label{1eq11.12}
\fl
& a_1=-\frac{1}{2}\left(e^x-e^y\right) + \frac{1}{8} \left(e^{2x}-e^{2y}\right), \qquad
a_2=\frac{1}{2}\left(e^x+e^y\right) -\frac{1}{4} \left(e^{2x}+e^{2y}\right), 
\nonumber \\
\fl
&a_3=-\frac{1}{2}\left(e^x-e^y\right) + \frac{3}{8} \left(e^{2x}-e^{2y}\right) +\frac{1}{12}\left(e^{3x}-e^{3y}\right) -\frac{1}{64}\left(e^{4x}-e^{4y}\right),
\quad {\rm etc}.
\end{eqnarray}
The exact Green function has the form
\begin{equation}
\label{1eq11.13}
G(x,y;k)=-\frac{i \pi}{2 \cos(i\pi k)}\left[\chi_+(x,k) + e^{\pi k} \chi_-(x,k)\right] \chi_-(y,k),
\end{equation}
where $\chi_\pm$ are defined in terms of the Bessel functions as
\begin{equation}
\label{1eq11.14}
\fl
\chi_\pm(x,k) \equiv \frac{i e^{x/2}}{\sqrt{2}}\left[J_{\nu_\pm}(-ie^x/2) + iJ_{-{\nu_\mp}}(-ie^x/2)\right],
\qquad \nu_\pm \equiv \frac{1}{2}\pm ik.
\end{equation}
Equation (\ref{1eq11.1}) with (\ref{1eq11.12}) gives the correct asymptotic expansion when $\arg k$ is fixed in  $0<\arg k<\pi$ (figure~2). 
\begin{figure}
\hspace{1cm}
\includegraphics[scale=0.7]{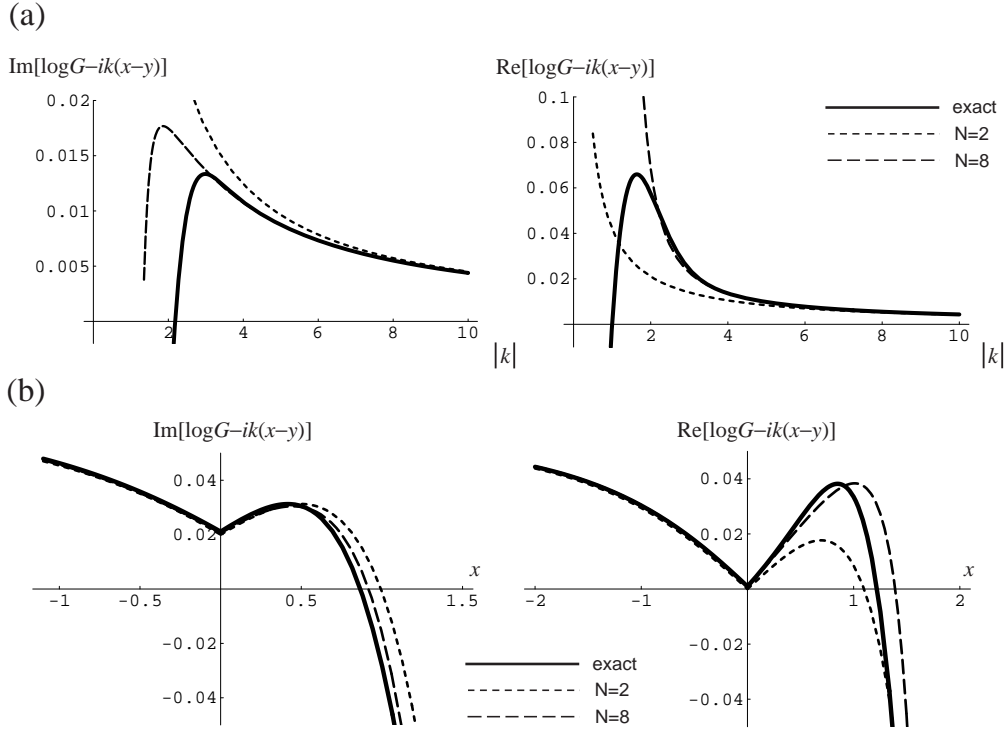}
\caption{
The imaginary and real parts of $\log G(x,y;k)-ik(x-y)$ for the potential $V(z)=e^z$ (example~3), 
(a) plotted as functions of $\vert k \vert$, with $\arg k$ fixed at $\pi/4$; $x=0.8$, $y=0$;
(b) plotted as functions of $x$, with $k=2.5 \exp(i\pi/4)$, $y=0$.  
Solid lines: the exact values.
Dashed lines: the expansion (\ref{1eq11.1}) to order $1/k^N$ ($N=2$ and $8$).
}
\end{figure}
However, this asymptotic expansion is not correct when $\vert k \vert \to \infty$ with fixed ${\rm Im}\,k$, irrespective of whether ${\rm Im}\,k>0$ or ${\rm Im}\,k=0$. The short-time expansion (\ref{1eq9.6}) with $g_n$ calculated from (\ref{1eq11.12}) is shown in figure~3. 
\begin{figure}
\hspace{1cm}
\includegraphics[scale=0.7]{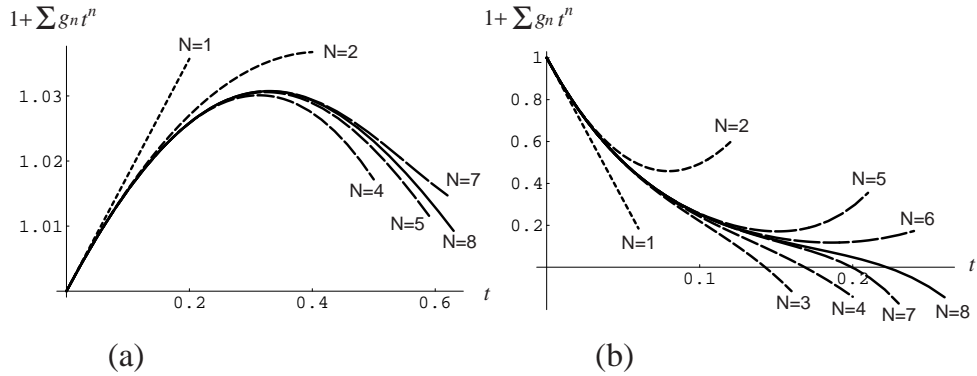}
\caption{
The series $1+\sum_{n=1}^N g_n t^n$ for the potential $V(z)=e^z$ (example~3),
plotted as a function of $t$ with various $N$. 
(a) $x=0.7$, $y=0$; (b) $x=3$, $y=0$.
}
\end{figure}

\bigskip
\noindent
{\bf Example 4.}\quad $V(z)=2 \log \cosh z, \quad f(z)=-\tanh z, 
\quad V_{\rm S}(z)=1-2\,{\rm sech}^2\, z.$

\nopagebreak
\smallskip
\noindent
This $V(x)$ tends to $+\infty$ linearly as $x\to \pm \infty$. Both $f(+\infty)$ and $f(-\infty)$ are finite, and conditions (iii) of section~7 are satisfied for any $N$.  In this case, we have
\begin{eqnarray}
\label{1eq11.15}
\fl
& a_1=x-y-2(\tanh x-\tanh y), \qquad \!
a_2=({\rm sech \,}x)^2+({\rm sech \,}y)^2 - (\tanh x)^2-(\tanh y)^2,
\nonumber \\
\fl
&a_3=-x+y + \frac{4}{3}(\tanh x- \tanh y)+ \frac{8}{3} [({\rm sech \,}x)^2 \tanh x-({\rm sech \,}y)^2 \tanh y], \quad {\rm etc}.
\end{eqnarray}
The exact $G$ for this potential is 
\begin{eqnarray}
\label{1eq11.16}
\fl
G(x,y;k)=\frac{k}{2 \pi \sqrt{k^2-1}}
\frac{\Gamma(\alpha)\Gamma(\frac{1}{2}-\alpha)\Gamma(\beta)\Gamma(\frac{1}{2}-\beta) \cos \alpha \pi \sin \beta \pi}
{\Gamma(\alpha-\beta)\Gamma(\beta-\alpha) \sin [(\beta-\alpha)\pi]}
 \frac{\eta_+(x) \eta_-(y)}{\cosh x \cosh y},
\end{eqnarray}
where
\begin{eqnarray}
\label{1eq11.17}
\fl
\eta_\pm(x,k)\equiv F\left(\textstyle{\alpha, \beta, \frac{1}{2}; -\sinh^2 x}\right) 
\nonumber \\ 
\mp 2 \frac{\Gamma(\frac{1}{2}+\alpha)\Gamma(1-\beta)}{\Gamma(\alpha)\Gamma(\frac{1}{2}-\beta)}
\sinh x F\left(\textstyle{\alpha+\frac{1}{2}, \beta+\frac{1}{2}, \frac{3}{2};-\sinh^2 x}\right),
\end{eqnarray}
with $\alpha \equiv \frac{1}{2}(-1-i \sqrt{k^2-1})$ and 
$\beta \equiv \frac{1}{2}(-1+i \sqrt{k^2-1})$. 
Here $F$ is the hypergeometric function defined by
$
F(\alpha,\beta,\gamma;z)
=\frac{\Gamma(\gamma)}{\Gamma(\alpha)\Gamma(\beta)}
\sum_{n=0}^\infty
\frac{\Gamma(\alpha+n)\Gamma(\beta+n)}{\Gamma(\gamma+n)}
\frac{1}{n!}z^n
$. 
Since $f(\pm \infty)$ are finite, expansion (\ref{1eq11.1}) with (\ref{1eq11.15}) is valid even when ${\rm Im}\,k=0$ (figure~4).
\begin{figure}
\hspace{1cm}
\includegraphics[scale=0.7]{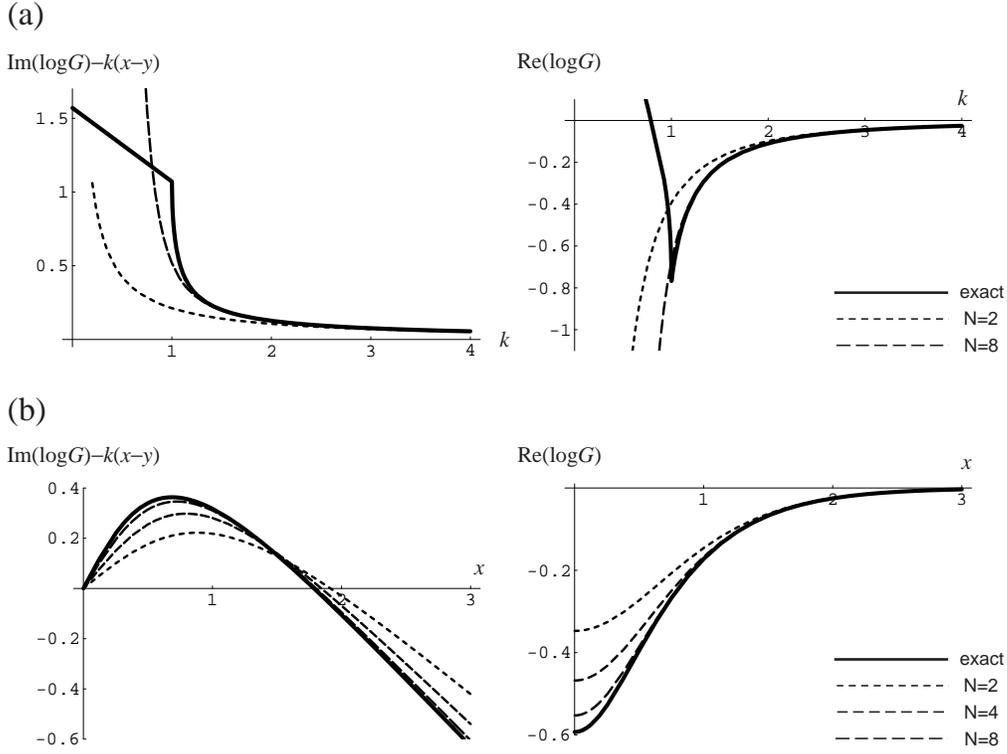}
\caption{
The imaginary and real parts of $\log G(x,y;k)-ik(x-y)$ for the potential $V(z)=2 \log \cosh z$ (example~4), (a) plotted as functions of real $k$, with $x=0.5$, $y=0$; 
(b) plotted as functions of $x$, with $k=1.2$, $y=0$. 
Solid lines: the exact values.
Dashed lines: the expansion (\ref{1eq11.1}) to order $1/k^N$.
(Since $k$ is real, $N=2,4,8$ are the same as $N=1,3,7$, respectively, for the imaginary part.)
}
\end{figure}
 As in example~1, the infinite series (\ref{1eq11.1}) is convergent for $\vert k \vert >1$.

\bigskip
\noindent
{\bf Example 5.}\quad $V(z)=2\vert z \vert, \quad f(z)=1-2\theta(z), 
\quad V_{\rm S}(z)=1-2\delta(z).$

\nopagebreak
\smallskip
\noindent
This is an example where $f(z)$ has a jump at $z=0$. 
This belongs to the case of (\ref{1eq7.1}) with $M=0$ and $C=-2$. 
For $y<0<x$, the exact Green function is
\begin{equation}
\label{1eq11.23}
G(x,y;k)=\frac{i+\sqrt{k^2-1}}{k}\exp\left[i\sqrt{k^2-1}(x-y)\right],
\end{equation}
and hence we have, as $\vert k \vert \to \infty$ ($0 \leq \arg k \leq \pi$),
\begin{equation}
\label{1eq11.24}
\fl
\log G(x,y;k)=ik(x-y)+\frac{1}{2ik}(x-y-2)-\frac{1}{(2ik)^3}\left(x-y-\frac{4}{3}\right)
+O(1/\vert k \vert^5).
\end{equation}
On the other hand, $a_1$ and $a_2$ are obtained from (\ref{2eq10.5}) as
\refstepcounter{equation}
\label{1eq11.25}
\addtocounter{equation}{-1}
\begin{equation}
\fl
a_1=\int_y^x\left[1-2\delta(z)\right]dz=x-y-2, \qquad
a_2=-1+2\delta(x)-1+2\delta(y)=-2.
\end{equation}
Comparing (\ref{1eq11.25}) with (\ref{1eq11.24}), we can see that $a_1$ is the correct coefficient of the expansion but $a_2$ is not. As shown in section~8, the coefficient $a_2$ needs to be corrected by $C^2/2$. Since $a_2+C^2/2=0$, we can see that (\ref{2eq8.11}) indeed agrees with (\ref{1eq11.24}).

\bigskip
\noindent
{\bf Example 6.}

\nopagebreak
\begin{equation*}
\fl
V(z)=
\cases{
e^z \\
1+z 
},
\qquad
f(z)=
\cases{
-e^z/2 \\
-1/2
},
\qquad
V_{\rm S}(z)=
\cases{
(e^{2z}/4)-(e^z/2)  & $(z<0)$ \\
1/4 & $(z>0)$
}.
\end{equation*}
\nopagebreak
This example belongs to the case of (\ref{1eq7.1}) with $M=1$ and $C=-1/2$. Now $f(z)$ is continuous and $f'(z)$ has a jump at $z=0$.  
For $y<0<x$, the exact $G$ has the form
\begin{equation}
\label{1eq11.26}
G(x,y;k)=\frac{-2\sqrt{2}ik e^{iKx}\chi_-(y,k)}{(K-i\nu_-)J_{\nu_-}(-i/2)+(iK+\nu_+)J_{-\nu_+}(-i/2)}
\end{equation}
with $K\equiv\sqrt{k^2-(1/4)}$, where $\chi_-$ and $\nu_\pm$ are defied by (\ref{1eq11.14}) . 
As shown in figure~5(a), we can check that (\ref{1eq6.5}) holds for $N\leq 3$ but not for $N\geq 4$.
\begin{figure}
\hspace{1cm}
\includegraphics[scale=0.7]{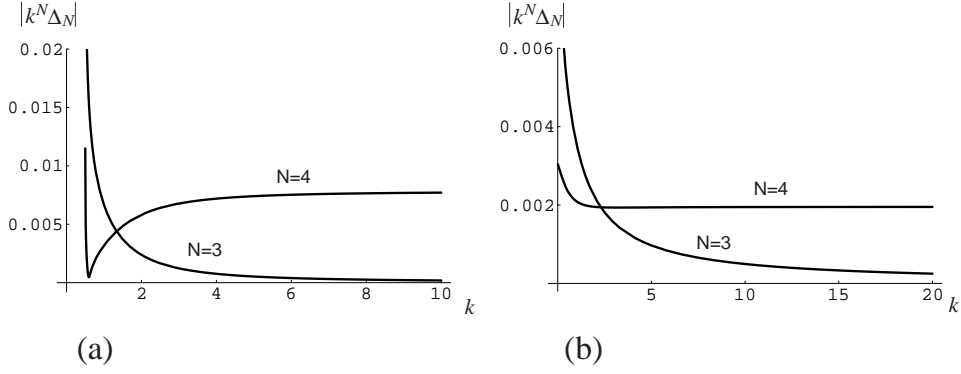}
\caption{
The graphs of $\vert k^N \Delta_N(x,y;k) \vert$ for examples~6 and 7, as functions of real $k$. ($N=3$ and $4$.)
(a) Example~6; $x=0.5$, $y=-0.5$. (b) Example~7; $x=0.5$, $y=-0.5$.
}
\end{figure}
It can be seen that the limit of $\vert k^4 \Delta_4\vert$ as $\vert k \vert \to \infty$ is $2^{-5}C^2\simeq0.0078$ as predicted by (\ref{2eq8.10}).
The coefficients of the expansion for $y<0<x$ are obtained from (\ref{1eq6.2}) as
\begin{eqnarray}
\label{1eq11.28}
a_1=\frac{1}{8}\left(-e^{2y}+4e^y+2x-3\right), \qquad 
a_2=-\frac{1}{4}\left(e^{2y}-2e^y+1\right),
\nonumber \\
a_3=\frac{1}{192}\left(3 e^{4y}-16e^{3y}-72e^{2y}+96e^y-12x-11\right),
\nonumber \\
a_4=\frac{1}{8}\left(e^{4y}-4e^{3y}-4e^{2y}+4e^y+1\right),
\quad {\rm etc}.
\end{eqnarray}
Since both $f(-\infty)$ and $f(+\infty)$ are finite, expansion (\ref{1eq11.1}) with (\ref{1eq11.28}) is correct to order $1/k^3$ for ${\rm Im}\,k\geq 0$. 
The correct expansion to order $1/k^4$ is obtained by adding $-C^2/2=-1/8$ to $a_4$ as in  (\ref{2eq8.11}). (See figure~6(a).)
\begin{figure}
\hspace{1cm}
\includegraphics[scale=0.7]{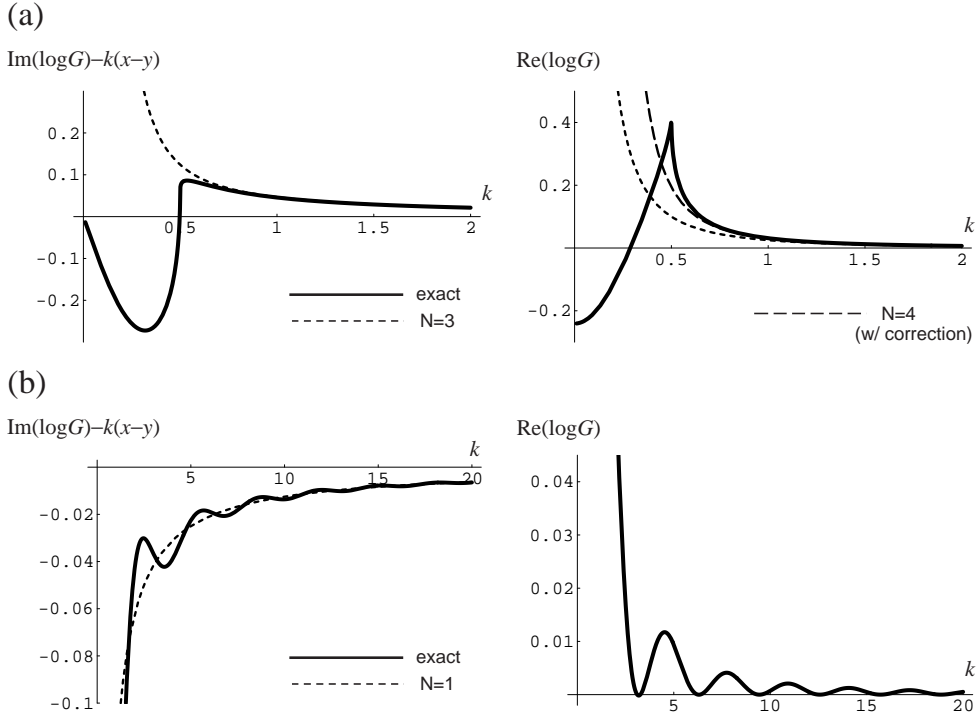}
\caption{The imaginary and real parts of $\log G(x,y;k)-ik(x-y)$ for the potential of example~6, plotted as functions of real $k$; 
(a) $x=0.5$, $y=-1$; (b) $x=2$, $y=1$. 
In (a), the dashed curve labeled ^^ ^^ $N=4$" shows the expansion to order $1/k^4$ with the correction term $-C^2/2$ added to $a_4$ as in equation (\ref{2eq8.11}). 
In (b), the envelope of the oscillation of ${\rm Re}(\log G)$ falls off like $1/k^2$.
}
\end{figure}

For $0<y<x$, the exact Green function  is
\begin{equation}
\label{1eq11.29}
G(x,y;k)=\frac{k}{K}
\left(e^{-iKy}
+\frac{A_-}{A_+}e^{iKy}
\right) e^{iKx},
\end{equation}
where
\begin{equation}
A_\pm\equiv
\left(i\pm\frac{\nu_-}{K}\right)J_{\nu_-}(-i/2)-
\left(1\mp i\frac{\nu_+}{K}\right)J_{-\nu_+}(-i/2).
\end{equation}
The $a_n$ obtained from (\ref{1eq6.2}) for $0<y<x$ are
\begin{equation}
\label{1eq11.31}
a_1=\frac{1}{4}(x-y), \qquad
a_2=-\frac{1}{2}, \qquad
a_3=-\frac{1}{16}(x-y), \quad {\rm etc}.
\end{equation}
We can show that the quantity $A_-/A_+$ is $O(1/\vert k\vert^2)$ as $\vert k \vert \to \infty$ ($0 \leq \arg k \leq \pi$).
Therefore, from (\ref{1eq11.29}) we can see that (\ref{1eq11.1}) with (\ref{1eq11.31}) is correct only up to order $1/k$ when ${\rm Im}\,k=0$ (figure~6(b)). For $\epsilon \leq \arg k\leq\pi-\epsilon$, this asymptotic expansion is correct to any order since $e^{iKy}$ in (\ref{1eq11.29}) vanishes faster than any power of $1/k$.

\bigskip
\noindent
{\bf Example 7.}\quad $V(z)=({\rm sgn}\,z)\left[\left(1+\vert z \vert\right)^{1/2}-1 \right], \quad f(z)=-\frac{1}{4}\left(1+\vert z \vert\right)^{-1/2}$,

\nopagebreak
$
\qquad \qquad \ V_{\rm S}(z)=\frac{1}{16} \left(1+\vert z \vert\right)^{-1}
+\frac{1}{8}\,({\rm sgn}\,z)\left(1+\vert z \vert\right)^{-3/2}.
$

\nopagebreak
\smallskip
\noindent
This is another case where $f'(z)$ has a finite jump at $z=0$. 
The exact Green function for $y<0<x$ can be expressed in terms of confluent hypergeometric functions as
\begin{equation}
G(x,y;k)=\frac{2ik \zeta_+(x,k)\zeta_-(y,k)}{\zeta'_+(0,k)\zeta_-(0,k)-\zeta_+(0,k)\zeta'_-(0,k)}
\end{equation}
where
\begin{eqnarray}
\fl
\zeta_\pm(x,k)
&\equiv
e^{ik(1\pm x)}\Biggl\{
q^{1/2}\Gamma(q)\!\left[
F\left(\textstyle{q,\frac{1}{2};-2ik(1\pm x)}\right)
\mp \frac{\sqrt{1\pm x}}{2}
F\left(\textstyle{q+1,\frac{3}{2};-2ik(1\pm x)}\right)
\right]
\nonumber \\
\fl
& \ \pm
\Gamma\left(\textstyle{q+\frac{1}{2}}\right)\left[
F\left(\textstyle{q+\frac{1}{2},\frac{1}{2};-2ik(1\pm x)}\right)
\mp \frac{\sqrt{1\pm x}}{2}
F\left(\textstyle{q+\frac{1}{2},\frac{3}{2};-2ik(1\pm x)}\right)
\right]
\Biggr\}
\nonumber \\
\fl
\end{eqnarray}
with $q\equiv i/(32k)$, and $\zeta_\pm'(x,k)\equiv (\partial/\partial x)\zeta_\pm(x,k)$. 
For $y<0<x$, the coefficients of the expansion calculated by our method are
\begin{eqnarray}
\label{1eq11.34}
\fl
a_1&=\frac{1}{16}\left[\log (1+x) +\log(1-y)\right]
-\frac{1}{4}\left[(1+x)^{-1/2}-(1-y)^{-1/2}\right],
\nonumber \\
\fl
a_2&=-\frac{1}{16}\left[(1+x)^{-1}+(1-y)^{-1}\right]
-\frac{1}{8}\left[(1+x)^{-3/2}-(1-y)^{-3/2}\right],
 \quad {\rm etc}.
\end{eqnarray}
This, too, is a case of (\ref{1eq7.1}) with $M=1$,
and so (\ref{1eq6.5}) holds for $N\leq 3$ but not for $N\geq 4$. Expansion (\ref{1eq11.1}) is now correct to order $1/k^3$ for ${\rm Im}\,k\geq 0$ (figure~7). We can see from figure~5(b) that $\vert k^4 \Delta_4\vert$ indeed approaches the predicted value $2^{-5}C^2 \simeq 0.0020$ as $\vert k \vert \to \infty$. (In this case, $C=1/4$.)
\begin{figure}
\hspace{1cm}
\includegraphics[scale=0.7]{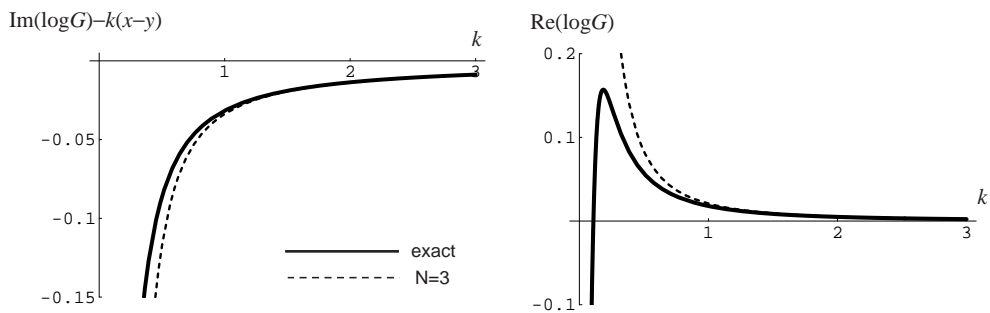}
\caption{The imaginary and real parts of $\log G(x,y;k)-ik(x-y)$ for the potential of example~7, plotted as functions of real $k$; $x=0.5$, $y=-0.5$.}
\end{figure}

\bigskip
\noindent
{\bf Example 8.}\quad $V_{\rm S}(z)=\vert z \vert$.

\nopagebreak
\smallskip
\noindent
Here we consider the case where $V_{\rm S}$ is given, and where $E_0 \neq 0$ (see section~10).
For $y<0<x$, the coefficients $a_n$ are obtained from (\ref{2eq10.5}) (or (\ref{1eq10.4}) and (\ref{1eq6.2})) as
\begin{eqnarray}
\label{2eq11.30}
\fl
a_1=\frac{1}{2}(x^2+y^2), \qquad
a_2=-x+y, \qquad
a_3=2-\frac{1}{3}(x^3-y^3), \qquad 
a_4=2(x^2+y^2), 
\nonumber \\
\fl
a_5=\frac{1}{2}(x^4+y^4)-5(x-y), \qquad
a_6=-\frac{16}{3}(x^3-y^3)+10, \quad {\rm etc.}
\end{eqnarray}
The exact Green function for $y<0<x$ is
\begin{equation}
G(x,y;k)=\frac{ik{\rm Ai}(x-k^2){\rm Ai}(-y-k^2)}{{\rm Ai}(-k^2){\rm Ai}'(-k^2)},
\end{equation}
where ${\rm Ai}(z)$ is the Airy function, and ${\rm Ai}'(z)$ is its derivative.
From the second equation of (\ref{1eq10.7}) we have 
$f(z)=(d/dz)\log {\rm Ai}(\vert z \vert -E_0)$, where $E_0$ is the smallest number satisfying ${\rm Ai}'(-E_0)=0$. (Numerically, $E_0\simeq 1.019$.) 
It is easy to see that $f''(0+)-f''(0-)=2$, so that this is a case of (\ref{1eq7.1}) with $M=2$, $C=2$. Expansion (\ref{1eq11.1}) with (\ref{2eq11.30}) is correct to order $1/k^5$ for ${\rm Im}\,k \geq \epsilon$. (Conditions (ii) of section~7 are satisfied since $f(z)$ behaves like $\vert z\vert^{1/2}$ as $\vert z \vert \to \infty$. This expansion is not valid for ${\rm Im}\,k=0$.) We can also check that the correct coefficient of order $1/k^6$ is not $a_6$ but $a_6+C^2/2=a_6+2$.

\section{Summary and remarks}
In this paper, we studied the high-energy asymptotic behavior of the Green function.
The expansion of $\log G$ in powers of $1/k$ (with $G$ defined by (\ref{1eq2.7}) and (\ref{1eq2.5})) is given by (\ref{1eq6.1}).
The coefficients of the expansion (equations (\ref{1eq6.2}) and (\ref{1eq4.3})) are expressed in terms of the coefficients $\tilde c_n$ for the expansion of the generalized reflection coefficient, which are calculated by using the formula (\ref{1eq3.8}) with (\ref{1eq3.2}). 
The remainder term $\Delta_N$ is also expressed in terms of $\tilde c_n$ (see (\ref{1eq6.3}), (\ref{1eq5.3}), (\ref{1eq4.12}), (\ref{1eq4.9}), (\ref{1eq4.6}) and (\ref{1eq3.9})).
Sufficient conditions for the validity of the expansion to order $N$ (equation (\ref{1eq6.5})) are given by (i), (ii), (iii) of section~7. These are not necessary conditions. Equation (\ref{1eq6.5}) holds under broader conditions as shown in section~8.

We assumed that the potential $V(x)$ is monotone for sufficiently large $\vert x \vert$, but this is not an essential restriction for our formalism. The formulas for the coefficients of the expansion and the remainder term are valid for any potential, as long as these quantities make sense. The particular shape of the potential is relevant only to the conditions for the validity of (\ref{1eq6.5}). 
The above mentioned assumption on the potential is used only in deriving the conditions for (\ref{1eq3.12}) quoted at the end of section~3.
Even for other kinds of potentials, we can use the same method to derive the criterion for (\ref{1eq6.5}).
For example, in this paper we excluded the cases where $V(x)$ oscillates indefinitely as $\vert x \vert \to \infty$, but our method can be applied to these cases as well, if only we study the validity of (\ref{1eq3.12}) (and its analogue for $R_l$) for such potentials in a similar way as in \cite{analysis}.

\appendix
\section{Properties of scattering coefficients for finite intervals}

Here we summarize some properties of the transmission and reflection coefficients for finite intervals.
(For details, see \cite{algebraic} and references therein.) 
First, it is obvious that 
\begin{equation}
\label{1eqa.1}
\tau(x,x;k)=1, \qquad R_r(x,x;k)=0, \qquad R_l(x,x;k)=0.
\end{equation} 

Let us assume that $f$ is piecewise smooth. 
We have, for finite $x$ and $y$,

\refstepcounter{equation}
\label{1eqa.2}
\addtocounter{equation}{-1}
\numpartsappendix
\begin{equation}
\tau(x,y;k)=e^{ik(x-y)}\left[1+O(1/\vert k \vert) \right],
\end{equation}
\begin{equation}
R_r(x,y;k)=O(1/\vert k \vert), \qquad R_l(x,y;k)=O(1/\vert k\vert),
\end{equation}
\endnumpartsappendix
as $\vert k \vert \to \infty$ ($0 \leq \arg k \leq \pi$).
The first equation of (\ref{1eqa.2}b) is a special case of (\ref{1eq3.3}) with $N=1$. 
For finite $x$ and $y$, equations (\ref{1eqa.2}) can be derived more directly from integral representations of the scattering coefficients.
 (See equations (1.14) of \cite{algebraic}. Alternatively, we can use equations (3.8) of \cite{algebraic} and the asymptotic forms of the functions $\alpha^\pm$ and $\beta^\pm$ to derive (\ref{1eqa.2}).)

From equations (3.5) and (3.8) of \cite{algebraic}, we can derive the differential equations
\refstepcounter{equation}
\label{1eqa.3}
\addtocounter{equation}{-1}
\numpartsappendix
\begin{equation}
\frac{\partial}{\partial x}\tau(x,y;k)=ik\tau(x,y;k)-f(x)R_r(x,y;k)\tau(x,y;k),
\end{equation}
\begin{equation}
\frac{\partial}{\partial y}\tau(x,y;k)=-ik\tau(x,y;k)-f(y)R_l(x,y;k)\tau(x,y;k),
\end{equation}
\begin{equation}
\frac{\partial}{\partial x}R_r(x,y;k)
=2ikR_r(x,y;k)+f(x)\left[1-R_r^2(x,y;k)\right],
\end{equation}
\begin{equation}
\frac{\partial}{\partial y}R_r(x,y;k)=-f(y)\tau^2(x,y;k),
\end{equation}
\begin{equation}
\frac{\partial}{\partial x}R_l(x,y;k)=-f(x)\tau^2(x,y;k),
\end{equation}
\begin{equation}
\frac{\partial}{\partial y}R_l(x,y;k)
=-2ikR_l(x,y;k)+f(y)\left[1-R_l^2(x,y;k)\right].
\end{equation}
\endnumpartsappendix
From (\ref{1eqa.2}) and (\ref{1eqa.3}) it follows that, as $\vert k \vert \to \infty$ ($0 \leq \arg k \leq \pi$), 
\refstepcounter{equation}
\label{1eqa.4}
\addtocounter{equation}{-1}
\numpartsappendix
\begin{equation}
\fl
\frac{\partial}{\partial z}\left[\frac{\tau(x,z)}{1+R_r(x,z)}\right]^2
=-2ik\left[\frac{\tau(x,z)}{1+R_r(x,z)}\right]^2 
+2f(z)e^{4ik(x-z)}
+ o(1),
\end{equation}
\begin{equation}
\fl
\frac{\partial}{\partial z}\left[\frac{\tau(z,x)}{1+R_l(z,x)}\right]^2
=2ik\left[\frac{\tau(z,x)}{1+R_l(z,x)}\right]^2 
+2f(z)e^{4ik(z-x)}
+ o(1),
\end{equation}
\endnumpartsappendix
\refstepcounter{equation}
\label{1eqa.5}
\addtocounter{equation}{-1}
\numpartsappendix
\begin{equation}
\fl
\int_a^b\left[\frac{\tau(z,c)}{1+R_r(z,c)}\right]^2dz
=
\frac{1}{2ik}\left\{
\left[\frac{\tau(b,c)}{1+R_r(b,c)}\right]^2-\left[\frac{\tau(a,c)}{1+R_r(a,c)}\right]^2
\right\}
+o(1/\vert k \vert),
\end{equation}
\begin{equation}
\fl
\int_a^b\left[\frac{\tau(c,z)}{1+R_l(c,z)}\right]^2dz
=
\frac{1}{2ik}\left\{
-\left[\frac{\tau(c,b)}{1+R_l(c,b)}\right]^2+\left[\frac{\tau(c,a)}{1+R_l(c,a)}\right]^2
\right\}
+o(1/\vert k \vert).
\end{equation}
\endnumpartsappendix

\section{Derivation of the short-time expansion}
Let $X\equiv x-y$. When (\ref{1eq9.1}) is substituted into (\ref{1eq9.4}), the integral of the $m$th-order term is proportional to
\begin{equation}
\label{1eqb.1}
\fl
\int_{-\infty}^\infty \frac{e^{i\kappa X}e^{-i\omega t}}{\kappa^{m+1}}d\omega
=-2i\frac{e^{-X^2/(4t)}}{\sqrt{t}}\int_{-\infty}^\infty\left(\frac{2t}{iX}\right)^m
\left(1+\frac{2\sqrt{t}}{iX}p\right)^{-m}e^{-p^2}dp,
\end{equation}
where we have changed the variable of integration from $\omega$ to $p\equiv (\kappa/\sqrt{t})-[iX/(t\sqrt{t})]$, and deformed the contour of integration. 
The right-hand side of (\ref{1eqb.1}) can be expanded in powers of $t$ by using the formula of Taylor expansion
\begin{equation}
\label{1eqb.2}
\fl
(1+\alpha)^{-m}=\sum_{j=0}^M \frac{(m+j-1)!}{j!\,(m-1)!}(-1)^j\alpha^j
+\frac{(M+m)!}{(M+1)!(m-1)!}
\frac{(-1)^{M+1}\alpha^{M+1}}{(1+\theta \alpha)^{M+m+1}},
\end{equation}
where $M$ is an arbitrary positive integer, and $0<\theta<1$. 
Applying (\ref{1eqb.2}) to (\ref{1eqb.1}), and carrying out the integration of each term, we have
\begin{eqnarray}
\label{1eqb.3}
\fl
\int_{-\infty}^\infty \frac{e^{i\kappa X}e^{-i\omega t}}{\kappa^{m+1}}d\omega
&=-2i\frac{e^{-X^2/(4t)}}{\sqrt{t}}
\Biggl\{
\sqrt{\pi}\sum_{j=0}^M\frac{2^{m+j}(2j+m-1)!}{(m-1)!(2j)!}\frac{t^{j+m}}{(iX)^{2j+m}}
\nonumber \\
\fl
&\quad + \frac{2^{m+2M+2}(2M+m+1)!}{(m-1)!(2M+2)!}\frac{t^{M+m+1}}{(iX)^{2M+m+2}}
\int_{-\infty}^\infty
\frac{p^{2M+2} e^{-p^2}dp}{\left[1+\frac{2\sqrt{t}}{iX}\theta p\right]^{2M+m+2}}
\Biggr\}.
\nonumber \\
\fl
\end{eqnarray}
From (\ref{1eq9.4}), (\ref{1eq9.1}) and (\ref{1eqb.3}), we obtain the expansion (\ref{1eq9.6}) with (\ref{1eq9.7}). 
The remainder term $\tilde \Delta_N$ can be written as
\begin{eqnarray}
\label{1eqb.4}
\fl
\tilde \Delta_N
&=\frac{1}{\sqrt{\pi}}\int_{-\infty}^\infty
\Delta'_N\biggl(x,y;\frac{p}{\sqrt{t}}+\frac{iX}{2t}\biggr)e^{-p^2}dp
\nonumber \\
\fl
& \quad+\frac{t^{N+1}}{\sqrt{\pi}}\sum_{m=1}^N\frac{b_m}{X^{2N-m+2}}
\frac{(-1)^{N+1}2^{2N-2m+2}(2N-m+1)!}{(m-1)!(2N-2m+2)!}
 \int_{-\infty}^\infty\frac{p^{2N-2m+2}e^{-p^2}dp}
{\left[1+ \frac{2\sqrt{t}}{iX}\theta_m p\right]^{2N-m+2}},
\nonumber \\
\fl
\end{eqnarray}
where $0<\theta_m<1$. 
The second term of (\ref{1eqb.4}) is obviously $O(t^{N+1})$ as $t \to 0$. 
Assuming that $X \neq 0$, the first term is $o(t^N)$ as $t \to 0$ if $\Delta'_N(x,y;k)=o(1/\vert k \vert^N)$ as $\vert k \vert \to \infty$ in the region $\epsilon \leq \arg k \leq \pi-\epsilon$.

\end{document}